\documentclass[12pt,a4paper,leqno]{amsart}
\usepackage{amsmath,amsthm,amssymb}

\newtheorem{theorem}{Theorem}
\newtheorem{example}{Example}
\newtheorem{proposition}[theorem]{Proposition}
\newcommand\bo[1]{{\bf #1}}

\newcommand\C{{\mathbb C}}

\newcommand\M{{\mathbb M}}
\newcommand\R{{\mathbb R}}

\newcommand\bT{{\mathbb T}}
\newcommand\bP{{\mathbb P}}
\newcommand\bE{{\mathbb E}}

\newcommand\cB{{\mathcal B}}
\newcommand\cD{{\mathcal D}}
\newcommand\cE{{\mathcal E}}
\newcommand\cF{{\mathcal F}}
\newcommand\cH{{\mathcal H}}
\newcommand\cL{{\mathcal L}}
\newcommand\cS{{\mathcal S}}
\newcommand\cP{{\mathcal P}}

\newcommand\cX{{\mathcal X}}

  \renewcommand\l{\lambda}
  
  \newcommand\s{\sigma}
  \newcommand\z{\zeta}
         
        \newcommand\e{\epsilon}
        \newcommand\om{\omega}

\newcommand\xchi{\smash{\raise 0.5ex\hbox{$\chi$}}}

\newcommand\bs[1]{{\boldsymbol #1}}

\renewcommand\vec[1]{\bs #1}
\renewcommand\Re[1]{\hbox{Re}(#1)}
\renewcommand\Im[1]{\hbox{Im}(#1)}

\begin{document}

\markboth{B.R.F. Jefferies}
{Hilbert's sixth problem}

% TOP MATTER

\title{ON HILBERT'S SIXTH PROBLEM} 
\author{B.R.F. Jefferies}
\address{School of Mathematics,\\ The University of New South Wales,\\ NSW 2052 AUSTRALIA}
\email{b.jefferies@unsw.edu.au}
\thanks{Dedicated to the memory of Igor Kluv\'anek.}

\begin{abstract}
Feynman path integrals are now a standard tool in quantum physics and
their use in differential geometry leads to new mathematical insights. A logical
treatment of quantum phenomena seems to require a sustained mathematical analysis of path integrals.
The subject is complicated by the fact that their application in flat space-time is quite different from how path integrals
are used in, say, topological quantum field theory, where there is no natural
notion of time translation. An historical background  is given in
this paper and a few mathematical approaches to Feynman path integrals in the context of
nonrelativistic quantum mechanics and scalar quantum fields with 
polynomial self-interactions are outlined.
\end{abstract}

\keywords{operator valued measure, Feynman path integral, functional integral, quantum mechanics}

\subjclass{Primary : 28B05, 46G10; Secondary 81S40, 81T08}

%\date{\today}

\maketitle

%\begin{history}
%\received{(Day Month Year)}
%\revised{(Day Month Year)}
%\end{history}

\section{Introduction}

David Hilbert presented his sixth problem at the Paris conference of the International Congress of Mathematicians, speaking on 8 August, 1900 in the Sorbonne \cite{H}. It roughly calls for the axiomatisation of physics. The explicit statement reads:
\medskip

\noindent 6. \textbf{Mathematical Treatment of the Axioms of Physics.} \textit{The investigations on the foundations of geometry suggest the problem: To treat in the same manner, by means of axioms, those physical sciences in which already today mathematics plays an important part; in the first rank are the theory of probabilities and mechanics.}\medskip

An overview of the mathematical state of Hilbert's sixth problem in 1976 was given by
A.S. Wightman \cite{Wight}. A different perspective is taken in this article in light of
the more recent interplay between physics and geometry, just in its infancy in Hilbert's time.

At the time that Hilbert stated his famous list of mathematical problems, Newtonian mechanics and Maxwell's equations were solving problems in the real world from building skyscrapers to transmitting voice over radio. There was not much work to be done writing down the axioms
of classical mechanics and fluid mechanics. Much had already been achieved by Newton, d'Alembert, Lagrange, Poisson, Hamilton, Stokes, Gibbs and many others. One specific mathematical problem Hilbert had in mind was the approach to equilibrium of Boltzmann's equation \cite{Wight}, to which C. Villani
recently made significant contributions \cite{Y}.

As in European politics of the time, there were rumblings in physics already. Maxwell's equations are invariant under Poincar\'e transformations, as
noticed by Minkowski in 1907 \cite{M1,M2}. The Michelson-Morley experiment seemed to indicate
strange viscous behaviour of the ether, governed by Lorentz transformations. But even Einstein's
special theory of relativity produced in his \textit{annus mirabilus} 1905\footnote{\texttt{http://en.wikipedia.org/wiki/Annus\_Mirabilis\_papers.}} can be reduced to a few simple principles: Einstein himself wrote them down.
Later, in 1915, during actual war, the General Theory of Relativity came in the form of simply formulated mathematical principles --- as long as you had differential geometry under your belt.

Einstein balked at the new quantum theory, but the quantum mechanics of nonrelativistic particles is also subject to mathematical axioms, essentially formulated by J. von Neumann \cite{vN} in 1932 using the then recent techniques of functional analysis and operator theory. Heisenberg's oracular `matrix mechanics' made sense after all. It seems that at the time of his formulation, he had not even {\it heard} of the concept of a matrix.
The combination of quantum physics and special relativity ran into immediate trouble with mysterious `infinities' appearing. Dirac's equation for an electron is relativistically invariant and gives an interpretation of the experimentally observed positron, but it is not an equation for a field, in the way that Maxwell's equations are. A proper description of subatomic particles must embody the wave-particle duality, so we must use field equations and these must be relativistically invariant.

The Nobel Prize in Physics 1965 was awarded jointly to Sin-Itiro Tomonaga, Julian Schwinger and Richard P. Feynman `for their fundamental work in quantum electrodynamics, with deep-ploughing consequences for the physics of elementary particles'. It is noteworthy that the prize was shared by physicists from formerly belligerent nations for their work done in the aftermath of WWII. 

The new theory of quantum electrodynamics was a stunning success. Measurements of the Lamb shift are accurate to within $10^{-8}$ parts, so that quantum electrodynamics is still
one of the most accurate physical theories constructed\footnote{\texttt{http://en.wikipedia.org/wiki/Precision\_tests\_of\_QED}}. Due to the physical irregularities of large clumps of matter, even Newtonian mechanics cannot be verified to this accuracy. General relativity deals with matter on the scale of planets, so accuracy to within a few percent is considered exceptional (although the equivalence of inertial and gravitational mass has been verified to within $10^{-12}$ parts \cite{Will}).

Feynman's great achievement was to develop a computational scheme for subtracting divergences from a putative perturbation series. He did this by `path integration'. Feynman diagrams describe terms in the perturbation series and Feynman devised the method by which to subtract divergences from the perturbation series, see \cite[Chapter 1]{CM} for an algebraic characterisation of Feynman diagrams. This is the stuff of current quantum field theory. Feynman path integrals are basic tools of theoretical physicists and mathematical physicists today. Feynman brilliantly conveyed the elementary nature of his physical thinking in a series of public lectures on light and matter in Auckland in 1979\footnote{\texttt{http://vega.org.uk/video/subseries/8}}.

E. Witten received the 1990 Fields Medal for his mathematical work on the interface between differential geometry and physics \cite{At} and also the inaugural 2012 Fundamental Physics Prize. He does masterly computations with Feynman path integrals, see for example \cite{W1}.
Unfortunately, the computations make no sense, but they produce the right answer, \textit{if you have enough insight}. It is not at all the same as computing an integral in the integral calculus of a function of one variable. No paid particle physicist understands all of the mathematics they use as a matter of course in their employ, because \textit{much of it makes no sense whatsoever}. 
The axiomatisation and proper understanding of relativistic quantum field theory seems a long way off.

The proper mathematical formulation
and solution of concrete physical problems
is a notoriously difficult endeavour. Recently,
bilinear integration, the approximation property for Banach spaces and
locally convex spaces have been used to
treat the connection between stationary state and time-dependent scattering
theory  \cite{GRJ},\cite{JefSBI}. Many hard analysts would consider such high-minded considerations, or even all of functional analysis itself, a useless abstraction. But there are many examples in history and mathematics where simple prejudice has impeded the advance of civilisation, see
\cite{Dys}, for example. The shelf life of mathematics is long. Euler probably did not imagine that
his work on number theory would feature in most 
electronic banking transactions in 2022. Grassmann did not foresee the
use of his work in particle physics.

It is well known that Feynman, a young theoretical physicist at the time, gave a talk at Cornell in 1948 at which the mathematician Mark Kac was in attendance \cite[p. 249]{Gleick}. Kac was able to see through all the witchcraft to deduce the connection between what Feynman was talking about with his path integral formulation of quantum mechanics and Brownian motion. The \textit{Feynman-Kac} formula is the fruit of this observation. 

The mathematical treatment of interacting systems satisfying the axioms of quantum field theory
is clearly a central issue for Hilbert's sixth problem \cite{Wight}.
The Osterwalder-Schrader axioms of {Euclidean quantum field theory} are famously verified in simple models in three and fewer space-time dimensions by employing the Feynman-Kac formula. The Wightman-G\aa rding axioms of relativistic quantum field theory are then verified by analytic continuation in time \cite{GJ}. This is surely one of Feynman's great mathematical legacies.

Nevertheless, the mathematical analysis of realistic problems
in quantum field theory remains a distant goal. Here is an excerpt from the description of one
of the millennium Clay prize problems \textit{Quantum Yang-Mills theory}, written by A. Jaffe and E. Witten.
% \smallskip

`Wightman and others have questioned for approximately fifty years whether
mathematically well-defined examples of relativistic, nonlinear quantum field theories
exist. We now have a partial answer: Extensive results on the existence and
physical properties of nonlinear QFTs have been proved through the emergence of
the body of work known as `constructive quantum field theory' (CQFT).
The answers are partial, for in most of these field theories one replaces the
Minkowski space-time $\M^4$ by a lower-dimensional space-time $\M^2$ or $\M^3$, or by a
compact approximation such as a torus. (Equivalently in the Euclidean formulation
one replaces Euclidean space-time $\R^4$ by $\R^2$ or $\R^3$.) Some results are known for
Yang-Mills theory on a 4-torus $\bT^4$ approximating $\R^4$, and, while the construction
is not complete, there is ample indication that known methods could be extended
to construct Yang-Mills theory on $\bT^4$.
In fact, at present we do not know any non-trivial relativistic field theory that
satisfies the Wightman (or any other reasonable) axioms in four dimensions. So even
having a detailed mathematical construction of Yang-Mills theory on a compact
space would represent a major breakthrough. Yet, even if this were accomplished,
no present ideas point the direction to establish the existence of a mass gap that is
uniform in the volume. Nor do present methods suggest how to obtain the existence
of the infinite volume limit $\bT^4 \to \R^4$.'

% \smallskip

The purpose of the present article is to suggest that a proper  understanding of Feynman's `mathematical' ideas may even lead to a complete solution of Hilbert's sixth problem, revolutionise geometry and low dimensional topology, make sense of string theory, elucidate scattering theory and prove the Riemann hypothesis---Hilbert's eighth problem \cite{H}. The last suggestion is not as far-fetched as it may seem, see, for example \cite[Chap. 2, \S 3, Chap. 4, \S 8]{CM} and \cite[\S 5.5]{Lap}. Nevertheless, mathematical research funding for the subject is hard to obtain. Perhaps Weierstrass would not have received research funding to eliminate infinitesimals from the calculus either. 
The historical analogy is spelt out by P. Cartier in \cite[\S VI]{CdWM} and \cite{Cart}.

Freeman Dyson, also a significant contributor to the theory of quantum electrodynamics, suggested  in his Gibbs Lecture `Missed Opportunities' \cite[\S 7]{Dys} that mathematicians should study Feynman path integrals. He was apparently not aware of the important contributions R.H. Cameron \cite{Cam,CS} and Yu. L. Daletskii \cite{Da}, although  E. Nelson \cite{Ne} is mentioned. Soon after Dyson's address, the lecture notes \cite{AHK} of S. Albeverio and R. H\o egh-Krohn appeared.

The monograph \cite{JL3} of G.W. Johnson and M. Lapidus appeared in 2000 and treats this material superbly at the level of a beginning graduate student. A lot of technical operator theory is covered.
 It runs to 792 pages, but still many interesting mathematical ideas concerning Feynman path integration
are omitted. It is not at all a subject that has been neglected by mathematicians.

There have been many different mathematical ways of dealing with Feynman's ideas,
so successful in modern differential geometry. The oscillatory integral approach of \cite{AHK} 
has been updated in \cite{AHKM}, where a more recent list of references may be found, see also 
\cite{AM,Maz1}. White noise calculus is exposed in \cite{HKPS} with an application
to Feynman path `integrals' or distributions in Chapter 12. Nevertheless,
\textit{functional integration} means different things to different 
mathematical physicists:
the monograph \cite{JL3} does not appear in the list of references in the second edition of
\cite{Si}, nor in \cite{Klau}.

Despite its wealth of material, the book \cite{JL3} does not treat what sort of `integration theory' Feynman was trying to use. In his 1948 paper \cite{Fe}, Feynman even writes about feeling like Cavalieri before the invention of the integral calculus, that is, the ideas he is proposing are mathematically \textit{ad hoc}. In his 1951 paper on operational methods, Feynman makes an
allusion to Clifford analysis\index{Clifford!analysis} techniques in \cite[Appendix B, p. 126]{Fe1}:
{\it The Pauli matrices (times $i$) are the basis for the
algebra of quaternions so that the solution of such
problems {\rm [concerning functional calculi]}
might open up the possibility of a true infinitesimal
calculus of quantities in the field of hypercomplex
numbers.}
He was essentially calling for the invention of quaternionic or Clifford analysis already developed by R. Fueter in 1935 \cite{Fue1,Fue2}: this is astonishing mathematical intuition for a nonmathematician.
Clifford analysis turns out to be the right tool
for the analysis of finite systems of noncommuting operators \cite{Jef}. V.P. Maslov developed the analysis of noncommuting systems of linear operators for applications to differential equations and asymptotic analysis
\cite{Mas,NSS}. A monograph further developing Feynman's operational methods has appeared \cite{JLN}.

The present article is  not meant to be a comprehensive survey of all mathematical
approaches to Feynman's path integral, although a number are mentioned.
I am confident that the references in the books and papers of the bibliography
cover a large chunk of the relevant mathematical literature, excluding the many
applications of the heuristic Feynman integral to topology and string theory.

The approach to path integrals here is motivated by the underlying integration theory,
that is, the necessity of integrating with respect to \textit{unbounded set functions} of the 
highly singular sort that arise in quantum physics. This is not a new idea. Stochastic
Integration is essentially integration with respect to the unbounded set function
associated with Brownian motion. Much of real variable harmonic analysis is
integration with respect to unbounded spectral measures on $L^p$-spaces $1 < p <\infty$, $p\neq 2$. Spectral shift functions, relevant to scattering theory, are analysed by integration
with respect to unbounded spectral set functions \cite{PS,PS1}.

The sort of integration theory that is in the background of the integrals described in \cite{JL3} is \textit{Riemann integration}. The Henstock-Kurzweil approach has also been considered \cite{Mul}. Because the Riemann integral is the backbone of undergraduate analysis, it is not worth making a song and dance over it in the context of Feynman path integrals.
The Henstock-Kurzweil integral has the attractive property that it may also be taught to first year university students.

When Henri Lebesgue published his modern theory of integration in 1904, British mathematicians, must have thought that this was another example of bizarre continental abstraction.\footnote{see J. Burkill, \textit{Biogr. Mems. Fell. R. Soc.} \textbf{24} (1978), 322-367.} But within thirty years, G.H. Hardy, J.E. Littlewood and R. Paley had used Lebesgue's theory of integration to lay the foundations of modern real variable harmonic analysis. At about the same time that von Neumann described the mathematical foundations of quantum theory, A.N. Kolmogorov used Lebesgue, Caratheodory and Hahn's theory to mathematically elucidate probability theory and statistics, in itself an important contribution to
the solution of Hilbert's sixth problem. It was not until after WWII that French abstraction had such an impact on analysis with L. Schwartz's \textit{Th\'eorie des distributions}.

\section{Feynman's path integral}\label{sec:Feynman}
It is worthwhile describing a few mathematical approaches to the Feynman path `integral' for a quantum particle moving in one space dimension. I will start with Feynman's description of his path integral which he developed by thinking about a remark of P. Dirac in the book \cite{Dirac} and in the paper \cite{Dirac1}.

For a quantum particle in one dimension, the evolution of its
state is expressed in terms of the continuous unitary group $e^{-itH/\hbar}$, $t\in\R$ of
operators, wher $\hbar = h/2\pi$ is Planck's constant $h$ divided by $2\pi$. The dynamics is specified by the Hamiltonian operator $H$, an observable. The
integral kernel $K(x,y,t)$, $x,y\in \R$, of the operator $e^{-itH/\hbar}$
is known as the {\it propagator} of the quantum system, although this may
only be a distributional expression. Knowledge of the propagator
therefore specifies the evolution of quantum states.

The {\it quantisation} of the dynamics of a classical mechanical
system means expressing the quantum propagator in terms of
classical quantities. Usually this is done by
replacing classical observables such as position and momentum
by their corresponding operator counterparts.
By contrast, R. Feynman \cite{Fe} tried to write
the quantum propagator $K(x,y,t)$ directly
in terms of the classical action
$S(q,\dot q,t)$ by means of the formula
\begin{equation}\label{eqn:1.1}
K(x,y,t) = N^{-1}\int_{C_{x,y}^{0,t}}e^{\frac{i}{\hbar}S(q,\dot q, t)}\,\cD q.
\end{equation}
Here $C_{x,y}^{0,t}$ is the set of all continuous paths $\omega$
such that $\omega(0)=x$ and $\omega(t) = y$, $N$ is a `normalisation factor'
and $\cD q$ is `uniform measure' on $C_{x,y}^{0,t}$.

For a single particle of mass $m$ on the line with a 
real valued potential $V$,
the action of a classical path $q:[0,t]\to\R$ is the expression
\begin{equation}\label{eqn:1.2}
S(q,\dot q, t) = \int_0^t \left[\frac{1}{2}m\, \dot q(s)^2 - V(q(s))\right]\,ds.
\end{equation}

An attempt can be made to
interpret formula (\ref{eqn:1.1}) by taking a positive integer $n$,
dividing the interval $[0,t]$
into $n$ equal parts and setting $K_n(x,y,t)$ equal to
\begin{align}\label{eqn:1.3}
&C_{n}\int_{\R^{n-1}}
\exp\left\{\frac{i}{\hbar}\sum_{j=1}^n\left[\frac{m}{2(t/n)}(x_j-x_{j-1})^2
-\frac{t}{n}V(x_j)\right]
\right\}\,dx_1\cdots
dx_{n-1},  
\end{align}
where $x_0=x$ and $x_n=y$ and the normalisation constant $C_n$ is given by
$$\left(\frac{-im}{2\pi\hbar(t/n)}\right)^{n/2}.$$
The exponent in the integrand of expression (\ref{eqn:1.3}) is $i/\hbar$ times
an approximation to the action (\ref{eqn:1.2}) in the case that $q$ is a polygonal path
\cite[Equation (7.4.1)]{JL3}. One might hope that the approximations $K_n(x,y,t)$,
$n=1,2,\dots$,
converge in an appropriate sense to the propagator $K(x,y,t)$.

One difficulty is that the integrand of expression (\ref{eqn:1.3})
has absolute value one, so the integral is not absolutely convergent.
Neverthless, $K_n(x,y,t)$ can be viewed as the kernel of
an operator which converges in the strong operator topology
as $n\to\infty$ for a large class of potentials $V$ \cite[Theorem 7.5.1]{JL3}.
The limiting operator is $e^{-itH/\hbar}$ for the associated quantum
Hamiltonian operator $H$. The reinterpretation of formula (\ref{eqn:1.1}) in terms of limits
of operator products was first given by E. Nelson \cite{Ne}
and is actually part of a more general theory of the approximation of semigroups
of operators \cite[Chapter 11]{JL3}; see Section \ref{sec:invsq} below. 

The
heuristic Feynman integral, taking the essential ideas of 
formula (\ref{eqn:1.1}), has produced useful formulae in low dimensional
topology and knot theory, see \cite[Section 20.2]{JL3} and \cite{Wi} for a
discussion of these developments. The geometric nature of these applications
means that there is no fundamental time parameter with which to
provide subdivisions and form operator products (but see the discussion of the
multiplicative property of the heuristic Feynman path integral on \cite[p. 648]{JL3}).
In the viewpoint of E. Witten \cite{Wi}, understanding four-dimensional
quantum gauge theory is intimately connected with understanding
Feynman path integrals over the space of connections on a four-manifold.
A. Connes hopes to treat quantum gravity by path integral techniques \cite[(1.892)]{CM}
and hence, the Riemann hypothesis \cite[Chapter 4, \S  8]{CM}.

Feynman arrived at formula (\ref{eqn:1.1}) by analogy. His `sum over histories' thinking in the context of light
is clearly conveyed in the videos of his Sir Douglas Robb lectures\footnote{\texttt{http://vega.org.uk/video/subseries/8}}. The Principle of Least Action is a formulation of classical mechanics that is equivalent to Hamiltonian mechanics. The quantisation of the relevant classical mechanical systems is clear. According to Feynman, formula (\ref{eqn:1.1}) is the quantum counterpart of the Principle of Least Action. The beauty of it is that taking the limit as $\hbar\to 0$, the stationary phase principle implies that the leading contributions to the integral come from the classical path, as may be proved for smooth potentials with suitably interpretion of the integral (\ref{eqn:1.1}).

As mentioned above, Mark Kac recognised (\ref{eqn:1.3}) from his own calculations with Wiener measure if $i/\hbar$ is replaced by $-1$. The Feynman-Kac formula is described in the operator setting in Section 3 below.
It is easily rewritten in terms of Brownian motion starting from a point in space.

Integration theory has come to mean the integration of
measurable functions defined on a set $\Omega$ with respect to an 
additive set function defined on a collection of subsets of $\Omega$.
The distinguishing feature of the set functions associated with
the Feynman integral of nonrelativistic quantum mechanics is that they
are {\it unbounded}\/ on the underlying algebra of cylinder sets,
even if they are analytically continued via a complex factor of time.

This was first noticed by R.H. Cameron \cite{Cam} and Yu. L. Daletskii \cite{Da}.
The calculation is as follows. Suppose
that
$t> 0$ and
$C^t$ is the collection of all continuous functions $\om:[0,t]\to\R$. Let $X_s(\om) =
\om(s)$ for every $\om\in C^t$ and
$0\le s
\le t$. Let $\phi,\psi\in L^2(\R)$ and $\l\in\C\setminus\{0\}$ with $\Re \l > 0$.
Then for every cylinder set
\begin{equation}\label{eqn:cylset}
E = \{X_{0}\in B_0,\ X_{t_1}\in B_1,\dots,X_{t_n}\in B_n,\ X_t\in B\ \}
\end{equation}
with $0 < t_1 <\cdots < t_n < t$, set
\begin{align}
&\mu^t_{\l,\phi,\psi}(E)\cr
&\quad=C_{n+1}\lambda^{\frac{d(n+1)}2}\int_{B}\int_{B_n}\cdots\int_{B_1}
\int_{B_0}\overline{\psi}(x_{n+1})
e^{-\frac{\lambda|x_{n+1}-x_n|^2}{2(t-t_{n})}}e^{-\frac{\lambda|x_n-x_{n-1}|^2}{2(t_n-t_{n-1})}}\cr 
&\qquad\qquad \cdots
e^{-\frac{\lambda|x_2-x_{1}|^2}{2(t_2-t_{1})}}
e^{-\frac{\lambda|x_{1}-x_0|^2}{2t_{1}}}\phi(x_0)\,dx_0dx_1\cdots
dx_ndx_{n+1}.\label{eqn:ints}
\end{align}
Here $C_{n+1} = (2\pi(t-t_n))^{-d/2}\dots(2\pi t_1)^{-d/2}$.
The integral (\ref{eqn:ints}), which is  associated with equation (\ref{eqn:1.3})
in the case that $V=0$,  is absolutely convergent because $\Re \l > 0$.

Then $\mu^t_{\l,\phi,\psi}$ defines an additive set function
on the algebra $\cS^t$ generated by all cylinder sets $E$ of the form (\ref{eqn:cylset})
as the times 
$0\le t_1 <\cdots < t_n\le t$ vary, the Borel subsets $B_0,\dots,B_n,B$
of $\R$ vary, and the index $n = 1,2,\dots$ varies. Here additivity means
that if $E\in \cS^t$ and $F\in\cS^t$ are disjoint sets, then
$$\mu^t_{\l,\phi,\psi}(E\cup F) = \mu^t_{\l,\phi,\psi}(E)+\mu^t_{\l,\phi,\psi}(F).$$
In the limiting case
with $\Re\l = 0$, the integrals (\ref{eqn:ints}) converge as improper iterated integrals.

Now fix the times $t_1,\dots,t_n$ and consider the algebra
$\cS^{t_1,\dots,t_n}$ generated by the cylinder sets of the form (\ref{eqn:cylset}), just as
the Borel subsets $B_0,\dots,B_n,B$ of $\R$ vary. A calculation shows that the total
variation $\|\mu^t_{\l,\phi,\psi}\|_{\cS^{t_1,\dots,t_n}}$ of
$\mu^t_{\l,\phi,\psi}$ over the algebra
$\cS^{t_1,\dots,t_n}$ is given by
\begin{align*}&\|\mu^t_{\l,\phi,\psi}\|_{\cS^{t_1,\dots,t_n}}\cr
&\quad=\left(\frac{|\l|}{\Re\l}\right)^\frac{d(n+1)}{2}\left(\frac{\Re\l}{2\pi
t}\right)^{d/2}\int_{\R}\int_{\R}
|\psi(x_{1})|e^{-\frac{\hbox{\tiny Re}\l|x_{1}-x_0|^2}{2t}}|\phi(x_0)|\,dx_0dx_1.
\end{align*}
The total variation $\|\mu^t_{\l,\phi,\psi}\|$ of $\mu^t_{\l,\phi,\psi}$
over the whole algebra $\cS^t$ is necessarily greater than the total variation
$\|\mu^t_{\l,\phi,\psi}\|_{\cS^{t_1,\dots,t_n}}$ for any choice of
times $0 < t_1 <\cdots < t_n < t$ and any $n = 1,2,\dots$. Hence, if $|\l| > \Re\l$
then either $\|\mu^t_{\l,\phi,\psi}\| = +\infty$, or $\|\mu^t_{\l,\phi,\psi}\| = 0$,
the last case occurring when either $\phi$ or $\psi$ is zero almost everywhere.
In the case $\Im\l \neq 0$, the additive set function $\mu^t_{\l,\phi,\psi}$
is highly singular.

\section{Evolution processes}
Rather than adopt Dirac's notation when considering path integrals,
I shall employ the concepts of direct interest in quantum modelling
of evolving systems:
the dynamical group $t\longmapsto e^{-itH}$, $t\in\R$, representing the evolution of quantum states and
the spectral measure $Q$ representing the observation of the position
of the system in the underlying classical configuration space $\Sigma$ of 
the system. If the system consists of $n$ interacting particles in three dimensional space, then 
$\Sigma=\R^{3n}$.

If $B$ is a Borel subset of $\Sigma$, then $Q(B)$ represents the
observable question as to whether or not the quantum system has its configuration or position in 
the set $B$.
For a quantum system in state $\psi$, the Hilbert space inner product
$0\le (Q(B)\psi,\psi)\le 1$ represents the probability that the configuration
or position
of the system belongs to the set $B$, see \cite{Mac}.
The analogous notions in classical mechanics give rise to a
measure or path integral concentrated on the collection of all classical paths of the
evolving system.

There is also an analogy with
the theory of time-homogeneous Markov processes $\langle X_t\rangle_{t\ge0}$ in probability theory, where
the normalised states are the initial probability distributions $\mu$, observables 
are represented by operators $\mu\longmapsto f.\mu$ for 
bounded Borel measurable functions $f$ and the distribution $P^\mu\circ X_t^{-1}$
of the process evolves as $e^{-tH}\mu$ for the generator $H$ of the process.

In Folland's mathematical treatment \cite{Fo2} of a corner of quantum electrodynamics,
many concessions are made to Dirac's notation. Here we are closer to Goethe's dictum:
\textit{Die Mathematiker sind eine Art Franzosen; redet man zu ihnen, so \'ubersetzen sie es in ihre Sprache, und dann ist es alsobald ganz etwas anders.}\footnote{Johann Wolfgang von Goethe, Maximen und Reflexionen, 2006 (Helmut Koopmann, ed.)}
(Mathematicians are like Frenchmen; if you talk to them, they translate it into their own language, and then it is immediately something quite different.)

We are led to the
following abstract setup described in \cite{J}.
Let $(\Sigma,\cE)$ be a measurable space. For each
$s \ge 0$, suppose that $\cS_s$ is a semi-algebra
of subsets of a nonempty set $\Omega$  (like intervals in $\R$) such that
$\cS_s\subseteq \cS_t$ for every $0\le s < t$.
For every $s\ge 0$, 
there are given functions $X_s:\Omega\to \Sigma$
with the property that $X_s^{-1}(B) \in \cS_t$
for all $0\le s\le t$ and $B\in\cE$. It follows that
the cylinder sets
\begin{align}\label{eqn:3.7}
E &= \{X_{t_1}\in B_1,\dots,X_{t_n}\in B_n\}\cr
& := \{\omega\in\Omega :  X_{t_1}(\omega)\in B_1,\dots,X_{t_n}(\omega)\in B_n\}\cr
&= X_{t_1}^{-1}(B_1)\cap\cdots\cap X_{t_n}^{-1}(B_n)
\end{align}
belong to $\cS_t$ for all 
$0 \le t_1 <\ldots <t_n \le t$ and $B_1,\dots,B_n\in\cE$.

Let $\cX$ be a Banach space. The vector space 
of all continuous linear operators $T:\cX\to\cX$ is denoted by $\cL(\cX)$.
It is equipped with the uniform operator topology, but measures with
values in $\cL(\cX)$ of practical interest are only $\s$-additive for the strong operator topology.

A {\it semigroup} $S$ of operators 
acting on $\cX$
is a map $S:[0,\infty)\to \cL(\cX)$ such that
$S(0) = Id_\cX$, the identity map on $\cX$ and
$S(t+s) = S(t)S(s)$ for all $s,t\ge 0$.
The semigroup $S$ represents the evolution of state vectors 
belonging to $\cX$.

It is an inconvenience of probability theory that $\cX$
should be the space of Borel measures with total variation norm
or the space of uniformly bounded Borel measurable functions with
the supremum norm, so that $S$ is not usually strongly
continuous. In this case, it is preferable that $\cX$ should carry
a topology weaker than the natural norm topology
for which well-posedness of the corresponding Cauchy problem is valid.

 An $\cL(\cX)$-valued {\it spectral measure}\/ $Q$ on $\cE$ is a map $Q:\cE\to \cL(\cX)$
that is $\s$-additive in the strong operator topology and satisfies
$Q(\Sigma) = Id_\cX$ and $Q(A\cap B) = Q(A)Q(B)$ for all $A,B\in\cE$.
In quantum theory, $Q$ is typically multiplication by characteristic
functions associated with the position observables, but the spectral measures
associated with momentum operators also appear.

Now suppose that $M^t:\cS_t\to \cL(\cX)$ is an additive operator valued
set function for each $t\ge 0$. The system
$$\left(\Omega,\langle \cS_t\rangle_{t\ge0},\langle M^t\rangle_{t\ge0};
\langle X_t\rangle_{t\ge0}\right)$$
is called a {\it time homogeneous Markov evolution process}\/
if there exists a $\cL(\cX)$-valued  spectral measure $Q$ on $\cE$
and a  semigroup $S$ of operators acting on $\cX$
such that for each $t\ge 0$, the operator $M^t(E)\in \cL(\cX)$ is given by
\begin{equation}\label{eqn:3.8}
M^t(E) =S(t-t_n)Q(B_n)S(t_n-t_{n-1})\cdots Q(B_1)S(t_1)
\end{equation}
for every cylinder set $E\in \cS_t$ of the form (\ref{eqn:3.7}) and the process
is called an {\it $(S,Q)$-process}\/.
The basic ingredients are the semigroup $S$ describing the evolution of states
and the spectral measure $Q$ describing observation of states
represented by vectors in $X$. An explicit proof 
that formula (\ref{eqn:3.8}) actually does define
an additive set function has been
written, for example, in
\cite[Proposition 7.1]{K2}.

In many significant examples the family $\cS_t$ is a
\textit{$\s$-algebra} of subsets of a nonempty set $\Omega$
and $M^t:\cS_t\to \cL(\cX)$ is \textit{$\s$-additive} for the strong operator
topology of $\cL(\cX)$, that is, the sum
$$M^t\left(\bigcup_{j=1}^\infty E_j\right)x=\sum_{j=1}^\infty M^t(E_j)x,$$
converges in the norm of $\cX$
for every $x\in \cX$ and for all pairwise disjoint $E_j\in\cS_t$, $j=1,2,\dots$\,. 

We have also accommodated the
examples arising in quantum physics where, as we have seen in Section \ref{sec:Feynman}, $\s$-additivity fails spectacularly.

\begin{example}\label{xmp:Markov} Let $(\Sigma,\cB)$ be a measurable space. Suppose that 
$$\left(\Omega,\cS,\langle P^x\rangle_{x\in\Sigma};\langle X_t\rangle_{t\ge0}\right)$$ 
is a Markov process with $P^x(X_0=x)=1$ for each $x\in\Sigma$. Then the \textit{transition functions}
$p_t(x,dy)$ are given by $p_t(x,B) = P^x(X_t\in B)$ for $x\in\Sigma$, $B\in \cB$ and $t\ge 0$.
For any signed measure $\mu:\cB\to\R$ and $t\ge 0$, the measure $S(t)\mu:\cB\to\R$ is given by
$$(S(t)\mu)(B)=\int_\Sigma p_t(x,B)\,d\mu(x),\quad\hbox{for }B\in\cB,$$
and the spectral measure $Q$ is given by $Q(B)\mu = \chi_B.\mu$, $B\in\cB$. Then $M^t(E)\mu$ defined by equation (\ref{eqn:3.8}), is the measure  $B\longmapsto \int_\Sigma P^x(\{X_t\in B\}\cap E)\,d\mu(x)$,
$B\in\cB$, so $M^t$ is an operator valued measure acting on the space $\cX$ of signed measures with the total variation norm. In general, the semigroup $S$ is not a C$_0$-semigroup on $\cX$. For a \textit{Feller process}, $S$ is a weak*-continuous semigroup of operators.

For the case of Brownian motion in $\R^d$, we have
$$p_t(x,dy)  =\frac{e^{-\frac{|x-y|^2}{2t}}}{(2\pi t)^{d/2}}\, dy,\quad t > 0,\ x\in \R^d.$$
\end{example}\medskip

In probability theory $\Omega$ is referred to as the \textit{sample space}. 
In quantum physics, where there is no underlying probability measure, $\Omega$ is referred to as the
\textit{path space}, where each $\omega\in\Omega$ is some function or \textit{path} $\omega:\R_+\to\Sigma$
and for each $t\ge 0$, the random variable $X_t:\Omega\to\Sigma$ is given by $X_t(\omega)=\omega(t)$
for every $\omega\in\Omega$. Here, a \textit{path integral} is a definite integral with respect to
the operator valued set function $M^t:\cS_t\to\cL(\cX)$, $t > 0$.

\begin{example} Let $(\Sigma,\cB,\mu)$ be a measure space and let $T_t:\Sigma\to\Sigma$, $t\in\R$,
be a group of measure preserving maps. Then $S(t):f\longmapsto f\circ T_{-t}$ for $f\in L^2(\mu)$ and $t\in\R$, defines a continuous unitary group $S$ of linear operators on $L^2(\mu)$. Observe that if $\delta_x$ is the unit point mass at $x\in\Sigma$, then 
$$\delta_x\circ T_{-t} = \delta_x\circ T_t^{-1} =  \delta_{T_tx}.$$

Suppose that the spectral measure $Q$ is defined by $Q(B)f = \chi_B.f$ for $B\in\cB$ and $f\in L^2(\mu)$. The measure $M^t$
defined by equation (\ref{eqn:3.8}) is given by 
$$M^t = S(t)(Q\circ\sigma^{-1}).$$
Here $\s:x\longmapsto \omega_x$  for $x\in\Sigma$. For each $x\in\Sigma$, the path $\omega_x$ is given by $\omega_x(t) = T_{t}x$, for $t\ge 0$.
The process
$\langle X_t\rangle_{t\ge0}$ is given by the evaluation maps
$X_t(\omega) = \omega(t)$ for $t\ge 0$ and $\omega\in\Omega= \s(\Sigma)$.

If we take $\Sigma$ to be the phase space of a system in classical mechanics with $\mu$ the 
Liouville measure, then the dynamical flow $t\longmapsto T_tx$, $t\in\R$, $x\in\Sigma$,
of the system satisfies the assumptions above. The operator valued path space measure $M^t$
acting on $L^2(\mu)$ is concentrated
on the sample space $\Omega$ of all classical paths in the time interval $[0,t]$.
\end{example}
Examples relevant to quantum mechanics are treated in greater detail the next section.
The viewpoint of operator valued measures, deals with the essential mathematical representations of the concepts of interest to physics:
the \textit{dynamical group} represented by $S$ and \textit{observations}, represented by the spectral measure $Q$. It is due to I. Kluv\'anek \cite{Kl,Kl1,Kl2,Kl3,K2}.

\section{The Feynman-Kac formula}\label{sec:FK}

Let $\Delta =\frac{\partial^2}{\partial x_1^2}+\dots+\frac{\partial^2}{\partial x_d^2}$ be the selfadjoint Laplacian in $L^2(\R^d)$ and set
$S_\lambda(t) = e^{t\Delta/(2\lambda)}$ for all $t\ge 0$ and $\lambda > 0$.
The exponential is defined by the functional calculus for selfadjoint operators.
Then for each $\lambda > 0$, the C$_0$-semigroup $S_\lambda$ is 
explicitly given by
\begin{equation}\label{eqn:3.1a}
(S_\lambda(t)\phi)(x) = \left(\frac{\lambda}{ 2\pi t}\right)^{d/2}\int_{\R^d}
e^{-\frac{\lambda}{ 2t}|x-y|^2}\phi(y)\,dy\quad a.e.
\end{equation}
for every $\phi\in L^2(\R^d)$ and $t > 0$. The same notation is used in the case that $\l$
is a complex number with $\Re\l \ge 0$ and $\l\neq 0$.
We use the notation $\C_+=\{z\in\C:\Re z \ge0\,\}$.
The integral (\ref{eqn:3.1a}) then converges in the mean square sense whenever $\l$ is a purely imaginary number.

Let
$Q(B):L^2(\R^d)\to L^2(\R^d)$ be the operator of multiplication by the
characteristic function of the Borel subset $B$ of $\R^d$. For a bounded
Borel measurable function $f$, the operator of multiplication by $f$ is written
as $Q(f)$. The operator $Q(f)$ is actually the integral $\int_{\R^d}f\,dQ$ of the 
bounded function
$f$ with respect to the spectral measure $Q:B\longmapsto Q(B)$.

There is a distinguished $(S_1,Q)$-process
$$\left(\Omega,\langle \cS_t\rangle_{t\ge0},\langle M^t_1\rangle_{t\ge0};
\langle X_t\rangle_{t\ge0}\right)$$
associated with the semigroup $S_1$ defined on $X=L^2(\R^d)$
by formula (\ref{eqn:3.1a}) in the case that $\lambda = 1$.
The space $\Omega$ of paths is the
set of all continuous functions from $\R_+$ into $\R^d$.
The random process $X$ is given by the evaluation
maps $X_s:\omega\longmapsto \omega(s)$
for all $\omega\in\Omega$ and $s\ge 0$ and
the semi-algebra $\cS_t$ is the $\s$-algebra generated by all 
cylinder sets $E$ of the form (\ref{eqn:3.7}) with times
$0 \le t_1 <\ldots <t_n \le t$ and $n=1,2,\dots$. 

For each $t > 0$, the operator valued set function $M_1^t:\cS_t\to \cL(L^2(\R^d))$
has the representation
\begin{equation}\label{eqn:WM}
\langle M_1^t(E)\psi,\varphi\rangle = \int_{{\R^d}}\left(\int_E \varphi(X_t(\om))\,dP^x(\om)\right)\psi(x)\,dx,\quad \quad E\in\cS_t,
\end{equation}
for every $\psi,\varphi\in L^2(\R^d)$,
with respect to \textit{Wiener measure} $P^x$ for $d$-dimensional Brownian motion starting at $x\in\R^d$. It is clear, then, that $M_1^t$ is an operator valued measure defined on the $\s$-algebra $\cS_t$ of subsets of $\Omega$.

The Feynman-Kac formula \cite[Theorem 12.1.1]{JL3} asserts that if $H := H_0\dot+ V$ is the form sum
of the free Hamiltonian $H_0$ and  the potential $V:\R^d\to\R$, then 
\begin{equation}\label{eqn:4.2}
e^{-tH} = \int_\Omega e^{-\int_0^tV\circ X_s\,d}\,dM^t_1.
\end{equation} Equation (\ref{eqn:4.2})
is therefore a reformulation of the Feynman-Kac formula in which the
operator $e^{-tH}$ is represented directly in terms of an integral
with respect to an operator valued measure. 

A similar argument still works if $-G$ is the generator of a C$_0$-semigroup $S$
and the $(S,Q)$-process
$\left(\Omega,\cS,\langle M^t\rangle_{t\ge0}, \langle \cS_t\rangle_{t\ge0};
\langle X_t\rangle_{t\ge0}\right)$ is $\s$-additive in the sense that
each set function $M_t$ is actually an operator valued measure, $\s$-additive
for the strong operator topology of $\cL(\cX)$. The equality
\begin{equation}\label{eqn:4.3}
S_V(t)=e^{-t(G+Q(V))} = \int_\Omega e^{-\int_0^tV\circ X_s\,ds}\,dM^t
\end{equation}
then holds for all bounded measurable potentials $V$ under weak measurability assumptions
on the process $X$. The equality holds in the limit if the left hand side
converges in the strong operator topology to a C$_0$-semigroup and the 
integrand converges to an $M^t$-integrable function for all $t > 0$
 as $V$ is approximated by bounded
cutoff potentials. Conditions for this to obtain are considered 
in \cite[Chapter 3]{J}.
The $(S_V,Q)$-process
$\left(\Omega,\langle \cS_t\rangle_{t\ge0},\langle M_V^t\rangle_{t\ge0};
\langle X_t\rangle_{t\ge0}\right)$
satisfies 
$$M_V^t = e^{-\int_0^tV\circ X_s\,ds}.M^t$$
 for every $t\ge 0$.
The integrands $e^{-\int_0^tV\circ X_s\,ds}:\Omega\to \R$, $t\ge0$, 
transforming one time homogeneous Markov evolution process into another
are known as \textit{multiplicative functionals} in probability theory.

Formulae of the type (\ref{eqn:4.3}) for operator valued measures
$M^t$ are also termed Feynman-Kac formulae
in \cite{J}, although the associated semigroup $S$ need not be
associated with the heat equation or any parabolic equation.
Formula (\ref{eqn:4.3}) suggests that we may
be able to write
\begin{equation}\label{eqn:4.4}
e^{-it(H_0+Q(V))} = \int_\Omega e^{-i\int_0^tV\circ X_s\,ds}\,dM^t_{-i} 
\end{equation}
for a suitable class of real valued potentials $V$, following the
idea of Feynman's formula (\ref{eqn:1.1}). However, as mentioned 
above, if $\cS_t$ denotes the
algebra generated by cylinder sets based before time $t$,
then the collection $\{M_{-i}^t(A) : A\in \cS_t\}$
of bounded linear operators is  unbounded in the operator norm
of $\cL(L^2(\R^d))$. Worse, the total variation of
the scalar set function 
$$( M_{\lambda}^t\phi,\psi):A\longmapsto  
( M_{\lambda}^t(A)\phi,\psi) = \mu^t_{\l,\phi,\psi}(A),\quad A\in\cS_t,$$
is either zero or infinite for $\phi,\psi\in L^2(\R^d)$ and
$\lambda\in\C_+$ with $\Im \lambda \neq 0$, so the right hand side of (\ref{eqn:4.4})
would have to be an integral with respect to a very singular object indeed.
Nevertheless, integration with respect to unbounded set functions is a hidden feature
of many problems in analysis --- a theme which is to be explored in a forthcoming monograph
\cite{CJO}.

\section{Finite path integrals}
Another of Feynman's ideas instigated quantum computing \cite{Fe2,Fe3}.
The state space here is a finite dimensional complex Hilbert space $\cH$ of dimension $n=1,2,\dots$,
in which the states of interest are \textit{qubits} $|q_1\dots q_n\rangle$ with $q_1,\dots,q_n\in \{0,1\}$.

The Hamiltonian $H:\cH\to\cH$ for an evolving system capable of quantum computations is represented by an hermitian matrix  with respect to some orthonormal basis of $\cH$, which may be identified with $\C^n$. For $1\le p < \infty$, let $\C^n_p$ be the space $\C^n$ with the $\ell^p$-norm
$$\|\vec x\|_p = \left(\sum_{j=1}^n|x_j|^p\right)^{\frac1p},\quad\vec x=(x_1,\dots,x_n)\in\C^n.$$
Then $\cH\equiv\C^n_2$. The space $\C^n$ with the norm $\|\vec x\|_\infty=\max_j|x_j|$
for $\vec x=(x_1,\dots,x_n)\in\C^n$ is written as $\C^n_\infty$.

The group $S(t) =e^{-itH}$, $t\in\R$, of matrices
has the property that for some unitary transformation $U:\cH\to\cH$,
the unitary matrix $US(t)U^*$ is diagonal and so is a contraction on $\C^n_\infty$ for every $t\in \R$. 

Let $Q:\cP(\{1,\dots,n\}) \to\cL(\cH)$ be the spectral measure defined by
$$Q(B)\vec x = \sum_{j\in B}x_j\vec e_j,\quad B\subseteq \{1,\dots, n\},\ \vec x = (x_1,\dots,x_n)\in \C^n.$$

It is not hard to see that there is an associated $(S,Q)$-process
$$\left(\Omega,\langle \cS_t\rangle_{t\ge0},\langle M^t\rangle_{t\ge0};
\langle X_t\rangle_{t\ge0}\right),$$
such that each set function $M^t:\cS_t\to\cL(\cH)$ is a matrix valued measure defined on a $\s$-algebra $\cS_t$. The set of paths $\Omega$ consists of
right-continuous jump functions $\om:\R_+\to \{1,\dots,n\}$ with finitely many jumps in each time interval \cite[Theorem 3]{Th} and $X_t(\om) = \om(t)$, $t\ge 0$.
There is an associated probabilistic Markov process
\begin{equation}\label{eqn:submarkov}
\left(\widetilde\Omega,\cS ,\langle {\widetilde P}^j\rangle_{j=1}^n;
\langle X_t\rangle_{t\ge0}\right)
\end{equation}
with $\widetilde\Omega$ consisting of paths with values in 
$\{1,\dots,n,\infty\,\}$ and a finite \textit{killing time}
$$\z(\om) =\inf\{t\in\R_+:\om(t)=\infty\,\},$$
see \cite[\S 9]{Th}.

Hamiltonians like $H$ also arise in \textit{network theory}, where ideas from
quantum physics are useful \cite{Ba}. In this case, we take the group
$S(t) =e^{-tH}$, $t\in\R$, of matrices and an associated $(S,Q)$-process
$$\left(\Omega,\langle \cS_t\rangle_{t\ge0},\langle M^t\rangle_{t\ge0};
\langle X_t\rangle_{t\ge0}\right).$$
Each set function $M^t:\cS_t\to\cL(\cH)$ is a matrix valued measure defined on a $\s$-algebra $\cS_t$. The set of paths $\Omega$ consists of
right-continuous jump functions $\om:\R_+\to \{1,\dots,n\}$ with finitely many jumps in each time interval  and $X_t(\om) = \om(t)$, $t\ge 0$. In these applications \cite{Ba}, the semigroup $S(t)$, $t\ge 0$, is actually a Markovian semigroup acting on $\cH \equiv \C^n_2$, so for each $k=1,\dots,n$, there exists a unique probability measure $P^k$
defined on the $\s$-algebra $\cS$ generated by $\cS_t$, $t\ge 0$, such that
$$M^t(E)\vec e_{k} =  \sum_{j=1}^nP^k(E\cap \{X_t=j\,\})\vec e_j,$$
for every cylinder set $E\in \cS_t$ of the form (\ref{eqn:3.7}) and every $t\ge 0$. Then
\begin{equation}\label{eqn:markov}
\left(\Omega,\cS ,\langle {P}^j\rangle_{j=1}^n;
\langle X_t\rangle_{t\ge0}\right)
\end{equation}
is a probabilistic Markov jump process with $P^j(X_0=j) =1$.

I cannot resist mentioning two of the many intriguing applications of quantum concepts to biology and chemistry pointed out by John Baez \cite{Ba}.

\begin{example}\rm In quantum chemistry, the states of a certain molecules like phosphorus pentachloride
can be represented by vertices on a \textit{Desargues graph}  $G$, which arises in projective geometry. For example, in the theory of \textit{bipartite Kneser graphs} $H(n,k )$, the
Desargues graph for iron pentacarbonyl is $H(5,2)$ \cite[Part 14]{Ba}.

Transitions between molecular states are represented by the edges in the graph $G$. The dynamics of an ion mutation is represented by Feynman's `sum over histories'---Feynman's path integral in this context is a nice $\s$-additive \textit{matrix valued measure} $M^t$ over paths in the Desargues graph $G$. The idea is lucidly explained in \cite[Part 14]{Ba}. By means of a \textit{Wick rotation} $t\longmapsto -it$, $t\in \R$, in the time parameter $t$, we obtain a random walk on the graph $G$, that is, a Markov jump process.

Writing $E\subset V\times V$ for the set of edges of the Desargues graph $G$ with vertex set $V$, then the Hamiltonian $H$ is given by
$$(H\psi)(x) =-\sum_{\{y:(x,y)\in E\, \}} \psi(y) +3\psi(x),\quad x\in V,$$
for $\psi\in L^2(V)$.
We're adding $3\psi(x)$ because there are 3 edges coming out of each vertex $x\in V$ in the relevant graph $H(5,2)$, so this is the rate at which the probability of staying at $x$ increases. The Hilbert space $L^2(V)$ is defined by the counting measure on the finite set $V$ and is identical to $\C^n_2$, if the Desargues graph $G$ has $n$ vertices.
Note that the Feynman integrals and diagrams \cite[Part 8]{Ba} can be calculated explicitly using classical measure theory in this example. Many application of random walks on
finite graphs may be found in \cite{BV}.
\end{example}

In the next application to biological systems, the state space $\cX$ is not finite dimensional
but the configuration space $\Sigma = \{0,1,2,\dots\,\}$ is discrete.

\begin{example}\rm Let $n=1,2,\dots$ and let $\cH$ be the Hilbert space of all entire functions $f:\C^n\to \C$ whose norms
$$\|f\| = \frac{1}{{(2\pi)}^\frac{n}2}\left(\int_{\C^n} e^{-|z|^2/2}|f(z)|^2\,d\mu(z)\right)^\frac12$$
are finite. Here $\mu$ is Lebesgue measure on $\C^n \sim \R^{2n}$. Then $\alpha_j:f \longmapsto \partial f/\partial z_j$, and $(\alpha_j^*f)(z) = z_jf(z)$ are the \textit{annihilation} and \textit{creation} operators on $\cH$,
respectively, for each $j=1,\dots, n$. The collection of monomials 
\begin{align*}
&z\longmapsto \frac{1}{\sqrt{{m_1!}\cdots  {m_n!}}}z_1^{m_1}\cdots z_n^{m_n}\\
& = \frac{1}{\sqrt{{m_1!}\cdots  {m_n!}}}\left((\alpha_1^*)^{m_1}\cdots (\alpha_n^*)^{m_n}1\right)(z),
\quad m_1,\dots,m_n = 0,1,2,\dots,
\end{align*}
is an orthonormal basis of $\cH$. The Hilbert space $\cH$ is associated with the 
\textit{Bargmann-Segal representation} in quantum mechanics, see \cite[pp. 39-50]{Fo}.

Suppose that $n=1$. An element $f\in\cH$ represents an estimate of the \textit{state} of a test tube full of amoebas
if   $f = \sum_{k=0}^\infty \psi_k{\mathbf z}^k$
is the power series of the entire function $f$ and $\psi_k$ is the probability of having $k$ amoebas in the test tube, $k=0,1,2,\dots$.

Then the \textit{Hamiltonian operator} $H$ representing amoeba evolution is given by  $-H = \alpha^*-1$. The unbounded selfadjoint operator $H$ describes the random division of amoebas
at a unit rate in time $t\ge 0$, because starting with one amoeba at time $t=0$, we have
$$(e^{-tH}1)(z) = e^{t(z-1)}= e^{-t}\sum_{k=0}^\infty \frac{t^k}{k!}z^k = \sum_{k=0}^\infty \psi_k(t)z^k,\quad z\in\C,$$
so that we obtain a Poisson process $\langle X_t\rangle_{t\ge 0}$ satisfying
$$\psi_k(t)=\bP(X_t=k)  = e^{-t}\frac{t^k}{k!},\quad k=0,1,\dots\,.$$

For $S(t)=e^{-tH}$, $t\ge0$, and the spectral
measure $B\longmapsto Q(B)$ of projections onto the orthonormal sets $\{{\mathbf z}^k/\sqrt{k!}:k\in B\}$ with $B\subset \{0,1,\dots\}$, applying the expression (\ref{eqn:3.8}) to complex a polynomial $p\in\cH$, we
obtain an $\cH$-valued measure $E\longmapsto M^t(E)p$, $E\in\cS_t$. 

The family $\cS_t$ may be taken as the $\s$-algebra generated by all cylinder sets $E$ of
the form (\ref{eqn:3.7}) for $t\ge 0$ fixed. The easiest way to see that
$M^t$ is a bona fide operator valued measure on $\cH$ is to observe that,  with respect to the orthonormal basis $\langle {\mathbf z}^k/\sqrt{k!}\rangle_{k=0}^\infty$, the Hilbert space $\cH$ is isometrically isomorphic to the 
sequence space $\ell^2$. Under this unitary map $U:\cH\to\ell^2$, the semigroup $e^{-tH}$, $t\ge 0$,
of operators on $\cH$ is associated with a positive semigroup $Ue^{-tH}U^*$
of contractions on $\ell^1$ and $\ell^\infty$ and interpolation gives the uniform boundedness
and $\s$-additivity of $UM^tU^*$, and so, of $M^t:\cS_t\to\cL(\cH)$.

In a predator-prey model \cite[Part 8]{Ba}, we have $n=2$ and $-H=\beta B+\gamma C +\delta D$
with $\beta,\gamma,\delta \ge 0$ and
\begin{align*}
B&= (\alpha_1^*)^2 \alpha_1-\alpha_1^*\alpha_1\\
C&=(\alpha_2^*)^2\alpha_1\alpha_2-\alpha_1^*\alpha_2^*\alpha_1\alpha_2\\
D&=\alpha_2-\alpha_2^*\alpha_2.
\end{align*}
Here the Feynman diagrams for the associated perturbation series correspond to
various predator-prey interactions.
\end{example} 

\section{Jump processes}\label{sec:jump}
The finite path integrals mentioned above are related to
the representation of solutions of evolution equations by
means of an integral over jump processes. M. Kac was the first
to notice in 1956 that hyperbolic 
equations can have a probabilistic representation \cite{Kac}.

Given a solution $(x,t)\longmapsto \varphi(x,t)$ of the wave equation
$$\frac{\partial^2\varphi}{\partial t^2} = v^2\Delta\varphi, \quad \partial_t\varphi(0,x)=0,\ x\in\R^d,$$
the function $(x,t)\longmapsto \bE(\varphi(x,\tau_t))$ is a solution of the damped equation
$$\frac{\partial^2\psi}{\partial t^2}+2a\frac{\partial\psi}{\partial t} = v^2\Delta\psi, \quad \psi(0,x)=\varphi(0,x),\ \partial_t\psi(0,x)=0,\ x\in\R^d,$$
for the process $\tau_t$, $t\ge0$, defined by
\begin{equation}\label{eqn:tau}
\tau_t=\int_0^t(-1)^{N_s}\,ds,\quad t\ge 0,
\end{equation}
for a Poisson process $N$ with intensity $a > 0$, that is,
$$\bP(N_s=k)  = e^{-as}\frac{(as)^k}{k!},\quad k=0,1,\dots,\ s\ge0,$$
and $\bE(f) = \int_\Omega f\,d\bP$ for a random variable $f$.

A systematic treatment of this phenomenon gave rise to
the analysis of \textit{random evolutions}, where there is a random
switching between modes of evolution \cite{KS}. The collection
of papers of reference \cite{Kac} gives an overview of the theory up 
to 1972 and R. Hersh himself  gives a brief personal history of the subject's
development in  \cite{Hersh}. In general, random evolutions may be viewed as
multiplicative operator functionals mapping one Markov evolution process
into another \cite[Chap. 5]{J}.

The Dirac equation in two space-time dimensions is another example where
there is a random switching involving a Poisson process between two energy states, see \cite{I2}, \cite{BJ2}, 
\cite[Chap 12]{CdWM1} and the references therein. The relevant $(S,Q)$-process
is a $\s$-additive operator valued measure acting on $L^2(\R,\C^2)$. The path space
is the set $\{x\pm \tau:x\in\R\,\}$ with the process $\tau$ given by formula (\ref{eqn:tau}) \cite[\S 7]{BJ2}.
Other types of representation by jump processes have been considered in
\cite{Com1,Com2,Com3,Com4} and \cite{Blan1,Blan2,Blan3,Blan4}.

It was pointed out in \cite[\S 6.3]{J} that, with 
the state space $\cX = L^p(\mu)$ for a measure $\mu$ and $1 \le p <\infty$, if $Q$ is 
the spectral measure of multiplication by characteristic functions and $S(t)=e^{tA}$, $t\ge 0$,
and the restriction of $A$ to $L^\infty\cap L^p(\mu)$ defines a weak*-continuous linear operator on $ L^\infty(\mu)$, then the associated $(S,Q)$-process is $\s$-additive and concentrated
on piecewise \textit{constant} paths. Finite path integrals correspond to the situation
where $\mu$ is counting measure on a finite set and $A$ is simply a matrix.

V.P. Maslov and A.M. Chebotarev treated such a situation 
in the papers \cite{MC1,MC2} by considering the Schr\"odinger equation in the momentum representation
with the potential $V$ as the Fourier transform of a complex Borel measure $\nu$ on $\R^d$.
With $Q$ being multiplication by characteristic functions and $S(t)= e^{-itV(i \nabla)}$, $t\ge 0$, acting on $L^2(\R^d)$, the operator $V(i\nabla)$
is convolution with respect to a complex measure and so it is weak*-continuous
on $L^\infty(\R^d)$ and the representation
$$e^{-it(H_0+V)}= \int_\Omega e^{-\frac{i}2\int_0^t|X_s|^2\,ds}\,d(\cF^{-1}M^t\cF)$$
is obtained on $L^2(\R^d)$. The Fourier-Plancherel transform 
$$(\cF\psi)(\xi)= \int_{\R^d}e^{-i\xi\cdot x}\psi(x)\,dx$$
 transforms $\psi\in L^2(\R^d)$ into the momentum representation.
The idea is refined further in \cite{Kolk1}.

\section{The attractive inverse square potential}\label{sec:invsq}
Suppose that $H_0$ is the selfadjoint extension of $\frac12\Delta$ in $L^2(\R^d)$
and $V:\R^d\to\R$ is a Borel measurable function.
In the paper \cite{Ne}, E. Nelson proved that
\begin{align}
e^{-it(H_0+Q(V))} &= \lim_{n\to\infty}\prod_{k=1}^n \left(e^{-itQ(V)/n}e^{-itH_0/n}\right)\cr
&= \lim_{n\to\infty}\int_\Omega e^{-i\sum_{j=1}^nV\circ X_{jt/n}t/n}\,dM^t_{-i},\label{eqn:5.1}
\end{align}
provided that the closure of $-i(H_0+Q(V))$ on the intersection
$\cD(H_0)\cap\cD(Q(V))$ of the domains of $H_0$ and $Q(V)$
is the generator of a C$_0$-semigroup on $L^2(\R^d)$. If 
$$F(\om) = f_1(\om(t_1))\cdots f_n(\om(t_n)),\quad \om\in\Omega,$$
is the product of uniformly bounded measurable functions
for $0\le t_1 < \cdots <t_n \le t$, taking separate limits of simple functions, we obtain
$$\int_\Omega F\,dM^t_{-i} =
e^{-i(t-t_n)H_0}Q(f_n)e^{-i(t_n-t_{{n-1}})H_0}\cdots Q(f_1)e^{-it_1H_0}.$$
All we use here is the separate $\s$-additivity of the expression
\begin{eqnarray*}
B_1\times \cdots\times B_n&\longmapsto&
M^t_{-i}(X_{t_1}\in B_1,\dots,X_{t_n}\in B_n)\\
&=& e^{-i(t-t_n)H_0}Q(B_n)e^{-i(t_n-t_{{n-1}})H_0}\cdots Q(B_1)e^{-it_1H_0}
\end{eqnarray*}
for $B_1,\dots,B_n\in\cB(\R^d)$. Now if we take $t_j = jt/n$ and
$$f_j(x)=e^{-i V(x)t/n},\quad x\in\R^d,$$
for $j=1,\dots, n$, then we obtain the second equality of equation (\ref{eqn:5.1}).

The first equality in (\ref{eqn:5.1}) is an instance of the 
\textit{Trotter product formula}, which has proved to be of wide use in probability theory and analysis, see \cite{JL3}.
Nelson takes the integral
$$\int_\Omega e^{-i\int_0^tV\circ X_s\,ds}\,dM^t_{-i} $$
to mean the limit 
$$\lim_{n\to\infty}\int_\Omega e^{-i\sum_{j=1}^nV\circ X_{jt/n}t/n}\,dM^t_{-i}$$
in the case that the potential
$V$ is sufficiently regular.
If $V$ is continuous off a set of capacity zero, then
$$\sum_{j=1}^nV\circ X_{jt/n}t/n \longrightarrow
\int_0^tV\circ X_s\,ds,\quad M^t_{1}\text{-- }a.e.\, .$$
The integral kernels of the operator products in equation (\ref{eqn:5.1}) are similar
to Feynman's approximation (\ref{eqn:1.3}).

In nonrelativistic quantum mechanics, the Coulomb potential 
$$V(x) = c/|x|,\quad x\in\R^3\setminus\{0\},$$
satisfies the conditions mentioned above for every $c\in\R$, but when $c > 1/8$, the \textit{attractive inverse square potential}
$$V_c(x) = -c/|x|^2,\quad x\in\R^3\setminus\{0\}$$ 
does not. In this case, the \textit{form sum}
of $H_0$ and $Q(V)$ is not even bounded below, indicating that the system may have negative infinite energy in certain states.

For $c\le 1/8$, the Feynman-Kac formula gives
\begin{equation}\label{eqn:5.2}
e^{-t(H_0/\l+iQ(V_c))} = \int_\Omega e^{-i\int_0^tV_c\circ X_s\,ds}\,dM^t_{\l},\quad 0< \l < (8c)^{-1}.
\end{equation}
Vitali's convergence theorem for analytic functions ensures that there exists a unique holomorphic operator valued function
$$\l\longmapsto \int_\Omega e^{-i\int_0^tV_c\circ X_s\,ds}\,dM^t_{\l},\quad \Re\l >0,\ \Im\l\neq 0,$$
with a continuous boundary values (\ref{eqn:5.2})
for all $0< \l < (8c)^{-1}$.

A calculation shows that there are also continuous boundary values
$$m\longmapsto \int_\Omega e^{-i\int_0^tV_c\circ X_s\,ds}\,dM^t_{-im},\quad m\neq 0$$
for every $c\in\R$, in particular, for $c > 1/8$, for which we write
$$
e^{-it(H_0/m\dotplus Q(V_c))} = \int_\Omega e^{-i\int_0^tV_c\circ X_s\,ds}\,dM^t_{-im}.$$
The calculation is made by looking at the resolvent $R_\l(\nu)$ of the semigroup
$$t\longmapsto \int_\Omega e^{-i\int_0^tV_c\circ X_s\,ds}\,dM^t_{\l},\quad t \ge 0,$$
and checking that $(\l,\nu)\longmapsto R_\l(\nu)$
is jointly holomorphic with continuous boundary values on the appropriate domain in $\C^2$. The kernel of the resolvent can be calculated explicitly in terms of Bessel functions \cite{Rad}.

For $c \le 1/8$, the operator $H_0\dotplus Q(V_c)$
defined by the right-hand boundary values of $\l  \longmapsto R_\l(\nu)$ 
coincides with the Friedrichs extension of 
the densely defined symmetric operator 
$H_0 + Q(V_c)$ \cite{Rad}.

However, for $c > 1/8$, the semigroup $t\longmapsto e^{-it(H_0\dotplus Q(V_c))}$, $t \ge 0$,  is a nonunitary continuous contraction semigroup. Nelson \cite{Ne} interprets the quantity 
$$1-\|e^{-it(H_0\dotplus Q(V_c))}\psi\|^2$$
as the probability that in the initial state $\psi$, the particle has collided with the centre of attraction by time $t > 0$. For $c \le 1/8$, the probability is zero because $H_0\dotplus Q(V_c)$ is selfadjoint.

The problem is unphysical, because a collision with the centre would ostensibly result in the creation of particle-antiparticle pairs, so a perturbative calculation in quantum field theory would be required. 

A similar phenomenon occurs with the Dirac equation in three space dimensions with an attractive Coulomb potential and an atomic number $Z$ greater than 137 and this is a more accurate model of a single electron cation theoretically collapsing under nuclear charge. The corresponding calculation in quantum field theory exhibiting the boundary conditions relating to particle-antiparticle pair production has been implemented. After imposing the appropriate boundary conditions, a selfadjoint Dirac Hamiltonian operator is obtained, so giving the dynamics of the evolving quantum system \cite{P}.

A Feynman path integral for the propagator of the Dirac operator  with radially symmetric potentials appears in \cite{J}. For the Coulomb potential, an atomic number $Z\le 137$ can be treated in this manner. An alternative path integral with respect to a countably additive measure is given in \cite{I3}. Here the Hamiltonian resolvent off the real line is represented as a path integral in the case that the Dirac operator is essentially selfadjoint. In both cases, the paths are continuous and piecewise smooth with values in the radial interval $(0,\infty)$.

\section{Feynman path integral for nonrelativistic quantum mechanics}\label{sec:Fpi}
 
Nelson's seminal paper \cite{Ne} and the work of R.H. Cameron and Storvick \cite{CS} suggests the following definition of the 
Feynman path integral relevant to the setting of nonrelativistic quantum mechanics \cite{J}. It corresponds to the `analytic in mass' integral treated in \cite{Ne}.

Let $\C_+=\{z\in\C:\Re z >0\}$.
Integration with respect to the family $\langle M_\lambda^t\rangle _{\lambda>0}$
of operator valued measures mentioned in Section \ref{sec:FK} above  may be used to control the convergence of
integrals with respect to the set functions $\langle M_\lambda^t\rangle
_{\lambda\in\C_+}$. First we have to make precise the idea
of integrating with respect to a {\it family}\/ of operator valued measures.

The space $L^1(\langle M_\lambda^t\rangle _{\lambda>0})$ of equivalence classes of functions integrable
with respect to each operator valued measure $M_\lambda^t$, $\lambda > 0$, is
equipped with a natural locally convex topology given by the
seminorms (\ref{eqn:9.0}) below, with respect to which it
is a sequentially complete locally convex Hausdorff topological space or, briefly, lcs. 
The seminorms defining the topology of $L^1(\langle M_\lambda^t\rangle
_{\lambda>0})$ are given by
\begin{equation}\label{eqn:9.0}
f\longmapsto \sup\left\{\int_{C^t} |f|\,d\mu^t_{\lambda,|\phi|,|\psi|}:\psi\in L^2(\R^d),\ \|\psi\|_2\le
1\right\}
\end{equation} 
for every $\phi\in L^2(\R^d)$ and $\lambda > 0$. The measures $\mu^t_{\lambda,|\phi|,|\psi|}$
are defined by equation (\ref{eqn:ints}).
Because $\mu^t_{\lambda,\phi,\psi} = ( M^t_\lambda\phi,\psi )$ for $\phi,\psi\in L^2(\R^d)$
and $\l >0$, we have
$$|( M^t_\lambda\phi,\psi )|  = \mu^t_{\lambda,|\phi|,|\psi|}.$$

The (sequential) completeness is a consequence of the operator valued measures $M_\lambda^t$ and
$M_\nu^t$ having disjoint support 
for all $\lambda>0$ and $\nu > 0$ such that $\lambda\neq
\nu$, that is, the operator valued measures live on
spaces of paths with distinct quadratic variation
according to a result of P. L\'evy. The measurability of functions belonging to
$L^1(\langle M_\lambda^t\rangle _{\lambda>0})$ is closely related
to the scale-invariant measurability
studied in \cite[Sections 4.2--4.4]{JL3}, except we do not require our
paths $\omega$ to satisfy $\omega(0) = 0$.

As is usual in integration theory, in order to
integrate with respect to the operator valued set functions
$M_\lambda^t:\cS_t\to\cL(L^2(\R^d))$ in the case that $\Im\lambda \neq 0$,
one starts with {\it simple functions}, in
this case, a finite linear combination $s = \sum_{j=1}^k c_k\chi_{E_j}$
of characteristic functions of sets $E_j\in \cS_t$, for $j=1,\dots,k$. Then linearity gives
$$\int_{C^t}s\,dM_\lambda^t = \sum_{j=1}^k c_kM^t_\lambda(E_k).$$
Let $\bo{{sim}}(\cS_t)$ be the linear space of simple functions $s$
based on $\cS_t$.

One idea is to give
the topology on $\bo{{sim}}(\cS_t)$
so that a net $\langle s_\alpha\rangle_{\alpha\in A}$ of simple functions converges to a
function $f$ if and only if it converges to $f$ in the quasicomplete space $L^1(\langle
M_\lambda^t\rangle _{\lambda>0})$ and the net is also Cauchy with respect to the seminorms
\begin{equation}\label{eqn:9.1}
p_{E,K,\phi}:s\longmapsto \sup_{\lambda\in K}\left\|\int_E
s(\omega)\,(M_\lambda^t\phi)(d\omega)\right\|_2
\end{equation}
as $E$ varies over cylinder sets
(\ref{eqn:cylset}), the function $\phi$ varies over $L^2(\R^d)$  and
$K$ varies over compact subsets of $\overline\C_+\setminus\{0\}$.
Then we can {\it define}
$$\int_E f(\omega)\,(M_\lambda^t\phi)(d\omega) := \lim_{\alpha\in A}\int_E
s_\alpha(\omega)\,(M_\lambda^t\phi)(d\omega)$$
so that the convergence is uniform  in the strong operator topology 
as $\lambda$ varies over
compact subsets of $\overline\C_+\setminus\{0\}$. It follows that the operator valued function
$$(E,\lambda)\longmapsto \int_E f(\omega)\,(M_\lambda^t\phi)(d\omega)$$
is finitely additive in $E\in\cS_t$ and analytic in $\lambda\in\C_+$. 
Because the convergence is uniform over compact subsets of 
$\overline\C_+\setminus\{0\}$, we obtain a continuous 
function for $\lambda\in\overline\C_+\setminus\{0\}$ and continuous
boundary values
$$\int_E f(\omega)\,(M_{is}^t\phi)(d\omega),\quad s \in  \R\setminus\{0\}.$$
Moreover, by construction, the space of integrable functions is sequentially complete
and quasicomplete in the given topology.

For the situation of interest---quantum mechanics---$\lambda$ is interpreted as $-i$
times a mass parameter $m$. It is not unreasonable to expect that the dynamics of a
quantum system should exhibit continuous dependence upon nonzero (and positive) mass.
Then analytic continuation in $\lambda$ from 
the boundary values on $(i\R)\setminus\{0\}$ to $\C_+$
can be achieved by the Poisson integral formula.

Of course, we could equally use $\lambda = -im$ with some smaller
interval $I$ of the mass parameter $m$, say, all positive real values.
Then we would look at analytic functions in $\C_+$ with continuous
boundary values on $-iI =\{-im:m\in I\,\}$.

By an approximating by simple functions, we quickly establish that
cylinder functions
$F:\omega\longmapsto f_1(\omega(t_1))\cdots f_n(\omega(t_n))$, $\omega\in \Omega$, with 
$f_1,\dots,f_n$ bounded and Borel measurable on $\R^d$ and $0\le t_1 <\dots <t_n\le t$
are integrable and
$$M^t_{-i} =  e^{-i(t-t_n)H_0}Q(f_n)e^{-i(t_n-t_{n-1})H_0}\cdots Q(f_1)e^{-it_1H_0}$$

The Trotter product formula helps to establish that the function
$$e^{-i\int_0^tV\circ X_s\,ds}:\omega \longmapsto e^{-i\int_0^tV(\omega(s))\,ds},
\quad \omega\in\Omega$$ 
with $V\in L^p(\R^d)+L^\infty(\R^d)$ is
integrable for $p > d/2$ and $d\ge 3$. Moreover, we have the representation
\begin{align}
e^{-itH(m)} &= \int_{C^t}e^{-i\int_0^tV\circ X_s\,ds}\,dM^t_\lambda \quad
\lambda = -im,\ m\in\R,\ m\neq 0,
\end{align}
relative to the operator $H(m) = -\Delta/(2m)+ V$. Here $X_s:\omega\longmapsto \omega(s)$
for all $\omega\in \Omega$ and $0\le s\le t$.

The notion of integrability just described is sufficent
to treat most potentials $V$ of physical interest
in quantum mechanics, although apparently not the attractive $1/r^2$-potential. 

Intimately connected with the
question of the {\it integrability}\/ of the
multiplicative functional 
$$e^{-i\int_0^tV\circ X_{s}\,ds},\quad t \ge 0,$$
with absolute value one is 
the {\it existence and uniqueness of the 
dynamics} of the quantum system associated with
the Hamiltonian `$H_0+Q(V)$'. Operator theoretic arguments feature 
heavily in the
proofs of integrability. 
In cases where $e^{-i\int_0^tV\circ X_{s}\,ds}$ is not
integrable or defined for all $t \ge 0$, modification of the
path space $\Omega$ and the set functions
$\langle M_\lambda^t\rangle
_{\lambda\in\C_+}$ can be made to reflect the
appropriate boundary conditions that lead to a
uniquely defined dynamical group, see for example \cite{BJ2}.
The boundary conditions need to be obtained from the physics.

In the next section, even the existence
of an appropriate multiplicative functional of
the process is an issue.

\section{Quantum field theory in Minkowski space}

The Feynman representation (\ref{eqn:1.1}) has an analogue in
quantum field theory discussed at the heuristic level
in \cite[Section 20.2]{JL3}, especially with regards
to knot theory and low dimensional topology. The Feynman-Kac formula
is also a tool in the construction of scalar quantum fields with 
polynomial self-interactions in Minkowski space with two and three space-time dimensions
\cite{GJ}. 

An overview of constructive quantum field theory 
relevant to the discussion below appears in \cite{Ja}.
A careful mathematical explanation of quantum field theory appears 
in G. Folland's book \cite[Chapter 6]{Fo2}, in which the author patiently explains
the mathematical difficulties associated with 
theoretical physics arguments in quantum electrodynamics. The experimentally
verifiable answers are obtained by employing 
superficially convincing mathematical analogies rather than deduction.
Our discussion here is limited to how one might make sense of
the path integral approach directly in Minkowski space. The book
\cite{GJ} treats \textit{Euclidean} quantum field theory and at the end of the day
applies the Wick rotation $t\longmapsto it$, $t\ge 0$.

The question arises of what is the evolution process
$$\left(\Omega,\langle \cS_t\rangle_{t\ge0},\langle M^t\rangle_{t\ge0};
\langle \Phi_t\rangle_{t\ge0}\right)$$
associated with a free quantum field and how can we represent
the dynamical group of an interacting field in the form
\begin{equation}\label{eqn:7.1}
e^{-itH} = \int_\Omega F_t\,dM^t,\quad t\ge 0,
\end{equation}
with
$t\longmapsto F_t,\ t\ge 0$, some multiplicative functional.
The multiplicative property 
$$F_{s+t}(\omega) = F_s(\theta_t\omega)F_t(\omega),\quad
\hbox{for  }\omega\in \Omega,\ s,t\ge 0$$
is just what's needed for the semigroup property of the integral.
The time shift transformation $\theta_t$ shifts the path or field by an amount $t$
in the time coordinate.
In the case of quantum mechanics, the multiplicative functional $t\longmapsto F_t$ is given by
$$F_t= e^{-i\int_0^tV\circ X_{s}\,ds},\quad t\ge 0,$$
and the potential $V$ is thought of as a \textit{perturbation}
of the free Hamiltonian $H_0$.

The situation is profoundly different in quantum field theory.
The Hamitonian operator $H$ is not constructed directly by
perturbation theory and the 
physically realistic multiplicative functional $F_t$
is not simply a Feynman-Kac functional $F_t = e^{-i\int_0^t V\circ \Phi_s\,ds}$---a
process of {\it renormalisation} is needed to construct $F_t$, $t\ge 0$.
See \cite{BJ2} for the renormalisation of the inverse potential in the Dirac equation on the line.

In this section, the evolution process
$\left(\Omega,\langle \cS_t\rangle_{t\ge0},\langle M^t\rangle_{t\ge0};
\langle X_t\rangle_{t\ge0}\right)$
associated with a free quantum field is constructed along the lines
suggested in \cite{GJ}. As in nonrelativistic quantum mechanics,
there is actually an associated family
$\left(\Omega,\langle \cS_t\rangle_{t\ge0},\langle M_\lambda^t\rangle_{t\ge0};
\langle X_t\rangle_{t\ge0}\right)$ of evolution processes
defined for all $\lambda
\in \overline\C_+\setminus\{0\}$, so that for each $t\ge 0$,
the equality
$ M^t=M_{-i}^t $ holds,
the function $\lambda \longmapsto M_\lambda^t(E)$, $\lambda\in\C_+\setminus\{0\}$, 
is  analytic and continuous in 
$\overline\C_+\setminus\{0\}$ for each $E\in \cS_t$, and
$M_\lambda^t$ is associated with the free Euclidean field
for each $\lambda > 0$, in the same way that the operator
valued measures defined in Section \ref{sec:FK} are associated with Wiener measure
by scaling. 

The section concludes with some comments about the
construction of the multiplicative functional $F_t$, $t\ge 0$,
for which the representation (\ref{eqn:7.1}) is valid,
although more work needs to be done  even in
two space-time dimensions, for which the Euclidean field theory
for polynomial self-interactions is well-understood.
\subsection{The free Euclidean field} The starting point
for $d$-dimensional Euclidean field theory is a probability measure $\mu$
defined on the Borel $\s$-algebra of the space $\cD'(\R^d)$
of Schwartz distributions defined on $\R^d$. The space of fields
$\phi\in \cD'(\R^d)$ plays the role of 
the paths $\omega \in C([0,\infty),\R^n)$ in quantum mechanics
and for the free field, $\mu$ is analogous to Wiener measure.

The inverse Fourier transform $S\mu:\cD(\R^d)\to \C$ of $\mu$ is defined by
\begin{equation}\label{eqn:7.2}
S\mu(f) = \int_{\cD'(\R^d)}e^{i\langle f,\phi\rangle}\,d\mu(\phi),\quad
f \in \cD(\R^d).
\end{equation}
In this section, the notation $\langle f,\phi\rangle = \phi(f)$
for $\phi\in \cD'(\R^d)$ and $f\in \cD(\R^d)$ is used to represent the
duality between $\cD(\R^d)$ and $\cD'(\R^d)$. The inner product of
a Hilbert space $\cH$ is written as $\langle\,\cdot\,,\,\cdot\,\rangle_{\cH}$.

The properties that the probability measure $\mu$ possesses
are formulated in terms of
the functional $S\mu$ defined on $\cD(\R^d)$ 
(the {\it Osterwalder-Schrader axioms}); these are listed
in \cite[pp. 89--90]{GJ}. Properties of the quantum field theory,
such as its time evolution
are derived from the the probability measure $\mu$. Because our
discussion is at the most basic level, we shall go directly
to the objects of interest for the free scalar quantum field
of mass $m=1$. 

For each $d = 1,2,\dots$, let $\Delta_d$ be the selfadjoint Laplacian in $L^2(\R^d)$
and let $H^{-1}(\R^d)$ be the Sobolev space of order
$-1$, defined as the completion of $L^2(\R^d)$ with respect
to the Hilbert space norm 
$f\longmapsto \langle(-\Delta_d +I)^{-1}f,f\rangle_{L^2(\R^d)}$,
$f\in L^2(\R^d)$. The Hilbert space $H^{-1}(\R^d)$
can be identified with tempered distributions $T\in \cS'(\R^d)$
for which 
$$\int_{\R^d}|\hat T(\xi)|^2(1+|\xi|^2)^{-1}\,d\xi<\infty.$$
The norm of $H^{-1}(\R^d)$ is denoted by
$\|\cdot\|_{H^{-1}(\R^d)}$ and the inner product by 
$\langle \,\cdot\,,\, \cdot\,\rangle_{H^{-1}(\R^d)}$.
The subscript is dropped from $\Delta_d$ if it is clear from the context.

Now suppose that $d \ge 2$. We take the free Euclidean field of unit mass to be the canonical Gaussian
process over $H^{-1}(\R^d)$, that is, a continuous linear map
$f\longmapsto \Phi_f$ from the Hilbert space $H^{-1}(\R^d)$
into the space $L^0(\Omega,\cF,\mu)$ of real random variables with respect to
a probability measure $\mu$ such that
\begin{equation}
\int_\Omega \Phi_f\Phi_g\,d\mu = 
\langle f,g\rangle_{H^{-1}(\R^d)},\quad f,g\in H^{-1}(\R^d).
\end{equation}

Just as Brownian motion is usually represented by Wiener measure on
the space of continuous functions, we take the probability measure $\mu$
to be the unique Gaussian measure defined on the Borel $\s$-algebra 
$\cF = \cB(\cD'(\R^d))$ of $\Omega = \cD'(\R^d)$ with mean zero
and variance $\int_\Omega |\phi(f)|^2\,d\mu(\phi) =\|f\|^2_{H^{-1}(\R^d)}$
for all $f\in \cD(\R^d)$. The measure $\mu$ exists by the Bochner-Minlos theorem 
\cite[Chap. IV Section 3]{GV}.

For any elements $f_1,\dots,f_n$ of $H^{-1}(\R^d)$, $n=1,2,\dots$,
the Borel probability measure
$$\mu\circ\left(\Phi_{f_1}\otimes\dots\otimes \Phi_{f_n}\right)^{-1} $$
is the unique mean zero Gaussian probability measure on $\R^n$,
whose covariance matrix is 
$$\left\{\langle f_j,f_k\rangle_{H^{-1}(\R^d)}\right\}_{j,k=1}^n.$$

Let $d=1,2,\dots$ and let $\cB_f(\R^d)$ denote the collection of all Borel sets of 
finite Lebesgue measure. Restricting the map $f\longmapsto \Phi_f$ to
the collection of all characteristic functions of sets of 
$A\in \cB_f(\R^d)\hookrightarrow L^2(\R^d)$,
the field may also be viewed as an $L^2(\mu)$-valued measure
$A\longmapsto \Phi_A$, $A\in \cB_f(\R^d)$, defined on
the $\delta$-ring $\cB_f(\R^d)$. In a similar way to the properties
of Brownian motion, the set of all $\omega\in \Omega$
for which there exists an open subset $U$ of $\R^d$
on which $f\longmapsto \langle f,\omega\rangle$, $f\in \cD(\R^d)$,
is a distribution of order zero, has $\mu$-measure zero \cite[Proposition 3.1]{CL}.

It is easy to see that every 
measurable function $f$ such that
$|f|\in H^{-1}(\R^d)$ is $\Phi$-integrable on $\R^d$ and
$\Phi_f = \int_{\R^d} f\,d\Phi.$ In particular, if $d\ge 2$ and $f\in L^2(\R^{d-1})$
and $B$ is a Borel subset of $\R$ with finite Lebesgue measure,
the function $f\otimes B:(x,t)\longmapsto f(x)\chi_B(t)$, $x\in \R^{d-1}$,
$t\in\R$, belongs to $H^{-1}(\R^d)$.

The following proposition is a slight reformulation of the
definition of sharp time fields.

\begin{proposition}\label{prp:7.1} The process $\Phi$ admits a continuous disintegration 
$X:\R\times \Omega\to \cS'(\R^{d-1})$ in the sense that
$X$ is $\cB(\R)\otimes\cF $-measurable and
\begin{enumerate}
\item[(i)] for each $\omega\in\Omega$, the $\cS'(\R^{d-1})$-valued function 
$t\longmapsto X_t(\omega) := X(t,\omega)$, $t\in \R$,
is continuous and 
\item[(ii)] for each $f\in \cS(\R^{d-1})$, the $L^2(\mu)$-valued function
$t\longmapsto \langle f, X_t(\,\cdot\,)\rangle$ is locally 
weakly Lebesgue integrable
in $L^2(\mu)$ and the equality
\begin{equation} 
\Phi_{f\otimes B} = \int_B \langle f, X_t(\,\cdot\,)\rangle\,dt, 
\end{equation}
holds in $L^2(\mu)$ for every $B\in \cB_f(\R)$.
\end{enumerate} 
Furthermore, $X$ is a version of the Ornstein-Uhlenbeck process, that is,
$X$ is Gaussian with mean zero and covariance
\begin{equation}\label{eqn:7.5}
\begin{split}
&\int_\Omega \langle f, X_s(\omega)\rangle
\langle g, X_t(\omega)\rangle\,d\mu(\omega)\\ &\qquad= 
\frac12\left\langle (-\Delta_{d-1} +I)^{-1/2}
e^{-|t-s|(-\Delta_{d-1} +I)^{-1/2}}f,g\right\rangle,
\end{split}
\end{equation}
for all $f,g\in \cS(\R^{d-1})$ and $ t,s\in\R.$
Any two such continuous disintegrations are indistinguishable.
\end{proposition}

It is possible to realise the process $X$ in a space much smaller that
$\cS'(\R^{d-1})$ \cite{Rock}.

If $\theta_s:\Omega\to\Omega$ denotes the time shift map
\begin{equation}\label{eqn:7.7a}
\langle f,\theta_s\omega\rangle = \langle \theta_{-s}f,\omega\rangle,\quad
\omega\in\Omega,\ f\in \cD(\R^{d}),
\end{equation}
 where
$\theta_{-s}f(x_1,\dots,x_d) = f(x_1\dots,x_{d-1},x_d-s)$,
then $X_t\circ\theta_s  = X_{t+s}$ for all $s,t\ge 0$.

The Hilbert space of the free field is $\cH = L^2(\mu\circ X_0^{-1})$.
By formula (\ref{eqn:7.5}), the image measure $\mu\circ X_0^{-1}$ is the unique
Gaussian measure on $\cS'(\R^{d-1})$ with mean zero and covariance
$(f,g)\longmapsto \frac12\left\langle 
(-\Delta_{d-1} +I)^{-1/2}f,g\right\rangle$, $f,g\in \cS'(\R^{d-1})$.
According to \cite[Corollary 6.2.8]{GJ}, the Hilbert space $\cH$
can be identified with the state space constructed from the
Osterwalder-Schrader axioms \cite[pp. 89--92]{GJ}.

The mapping $f\longmapsto f\circ X_0$, $f\in \cH$,
is an isometry between $\cH$ and $L^2(\Omega,\cF_0,\mu|_{\cF_0})$,
where $\cF_0$ is the $\s$-algebra generated by the random
variables 
$$\{\Phi_f:f\in H^{-1}(\R^d),\ \hbox{supp}\,f\subset \R^{d-1}\times\{0\}\ \}.$$
In particular, for each $u\in \cH$ and $t\ge 0$, there exists a unique
element $e^{-tH_0}u\in \cH$ such that
$$(e^{-tH_0}u)\circ X_0 = E(u\circ X_t|\cF_0),$$
where the right-hand side of the equation is the
the $\mu$-conditional expectation of $u\circ X_t$ with respect to $\cF_0$.
The Markov property of the process $\langle X_t\rangle_{t\ge0}$ ensures that
$t\longmapsto e^{-tH_0}$, $t\ge 0$,
is a Markov semigroup acting on $\cH$ and its generator $H_0$
is a positive selfadjoint operator---the free Hamiltonian
of the quantum field \cite[Section 5.1]{AHK1}. The 
general proof
given in \cite[Theorem 6.1.3]{GJ} for the construction
of the Hamiltonian operator avoids the use of sharp-time fields,
whose existence is problematic in more general situations.

\subsection{Evolution processes associated with the free field}
Let $\cH$ be the Hilbert space of Section 7.1 and $H_0$ the free
Hamiltonian defined in $\cH$. The spectral measure $Q$ of multiplication 
by characteristic functions of Borel subsets of $\cS'(\R^{d-1})$ acts on
$\cH$. It is the spectral measure associated with the position
operators $q(f)$, $f\in \cS(\R^{d-1})$, of the quantum field 
mapping $F\in \cH$ to the function $\phi\longmapsto \phi(f)F(\phi)$
, $\phi\in  \cS'(\R^{d-1})$, that is, for each $f\in \cS(\R^{d-1})$, 
the spectral measure of the selfadjoint operator $q(f)$
is $Q\circ \langle f,\cdot\rangle^{-1}$.

As in the consideration of quantum mechanics in Section 3,
we set
\begin{equation}\label{eqn:7.8}
S_\lambda(t) = e^{-(t/\lambda)H_0},\quad t\ge 0,\ \lambda \in \overline\C_+\setminus\{0\}.
\end{equation}
The operator is defined by the functional calculus for
selfadjoint operators.

Let $\langle X_t\rangle_{t\ge0}$ be the Ornstein-Uhlenbeck process defined
in Proposition \ref{prp:7.1} and let $\cS_t$ be the algebra
defined by all cylinder sets
\begin{equation}\label{eqn:7.9}
E = \{ X_{t_1}\in B_1,\dots,X_{t_n}\in B_n \}
\end{equation}
for $0 \le t_1 <\cdots <t_n \le t$, $B_1,\dots,B_n\in \cB(\cS(\R^{d-1}))$ and $n=1,2,\dots$\,.
Then for each $\lambda \in \overline\C_+\setminus\{0\}$, we have
an associated $(S_\lambda,Q)$-process
$$\left(\Omega,\langle \cS_t\rangle_{t\ge0},\langle M_\lambda^t\rangle_{t\ge0};
\langle X_t\rangle_{t\ge0}\right).$$

For $\lambda > 0$ and $E$ defined by (\ref{eqn:7.9}), the Markov property
for $\langle X_t\rangle_{t\ge0}$ ensures that
\begin{equation}\label{eqn:7.10}
\langle M_\lambda^t(E)u,v\rangle = 
\mu(\overline{v}\circ X_{t/\lambda}
\{ X_{t_1/\lambda}\in B_1,\dots,X_{t_n/\lambda}\in B_n \}
u\circ X_{0}),\ u,v\in\cH,
\end{equation}
so by analytic continuation from $\lambda > 0$, the operator
valued set functions $\langle M_\lambda^t\rangle_{t\ge0}$
are defined independently of the version of 
$\langle X_t\rangle_{t\ge0}$ chosen for every 
$\lambda \in \overline\C_+\setminus\{0\}$.

It follows from the representation (\ref{eqn:7.10})
that $M_\lambda^t$ is the restriction to $\cS_t$
of an operator valued measure, denoted again by the same symbol,
for each $\lambda > 0$.

Following the argument of Section 9 for the case of quantum mechanics,
a net $\langle s_\alpha\rangle_{\alpha\in A}$ of simple functions converges to a
function $f$ if and only if it converges to $f$ in the quasicomplete space $L^1(\langle
M_\lambda^t\rangle _{\lambda>0})$ and the net is also Cauchy with respect to the seminorms
\begin{equation}\label{eqn:7.11}
p_{E,K,\phi}:s\longmapsto \sup_{\lambda\in K}\left\|\int_E
s(\omega)\,(M_\lambda^t\phi)(d\omega)\right\|_\cH
\end{equation}
as $E$ varies over cylinder sets
(\ref{eqn:7.9}), the function $\phi$ varies over $\cH$  and
$K$ varies over compact subsets of $\overline{\C}_+\setminus\{0\}$.
Then we can {\it define}
$$\int_E f(\omega)\,(M_\lambda^t\phi)(d\omega) := \lim_{\alpha\in A}\int_E
s_\alpha(\omega)\,(M_\lambda^t\phi)(d\omega)$$
so that the convergence is uniform  in the strong operator topology 
as $\lambda$ varies over
compact subsets of $\overline{\C}_+\setminus\{0\}$. It follows that the operator valued function
$$(E,\lambda)\longmapsto \int_E f(\omega)\,(M_\lambda^t\phi)(d\omega)$$
is additive in $E\in\cS_t$, analytic in $\lambda\in\C_+$
and continuous on $\overline{\C}_+\setminus\{0\}$.
The appropriate notions of measurability and null sets here is with respect
to the family $\langle M_\lambda^t\rangle _{\lambda>0}$ of
operator valued measures.

\subsection{Multiplicative functionals of the free field and renormalisation}
In the present context, to say that
$F_t$, $t\ge 0$, is a multiplicative functional means that
for each
$t\ge 0$, the function 
$F_t:\Omega\to\C$ is  measurable with respect to the $\sigma$-algebra
$\sigma(\cS_t)$ (or perhaps, an appropriate completion) and 
\begin{equation}\label{eqn:7.13}
F_{s+t} =
F_t\circ\theta_sF_s,\quad
\langle M_\lambda^{t+s}\rangle _{\lambda>0}\hbox{-a.e.},
\end{equation}
 where $\theta_s:\Omega\to\Omega$ is the shift map given by
formula (\ref{eqn:7.7a}) for all $s\ge0$ and $\omega\in\Omega$. 

If $F_t$, $t\ge 0$, is a multiplicative functional which is
integrable in the sense of Section \ref{sec:Fpi}, then the operators
$$S_\lambda^F(t) = \int_\Omega F_t\,dM_\lambda^{t},\quad t\ge 0,$$
have the semigroup property for all $\lambda\in\overline{\C}_+\setminus\{0\}$
because the equality 
$$\int_\Omega F_t\circ\theta_sF_s\,dM_\lambda^{t+s} = 
\int_\Omega F_t\,dM_\lambda^{t}\int_\Omega F_s\,dM_\lambda^{s}$$
holds for all $\lambda > 0$, and so for all $\lambda\in\overline{\C}_+\setminus\{0\}$
by analytic continuation. We are seeking an integrable muliplicative
functional $F_t:\Omega \to \bT$ (with $\bT$ the unit circle in $\C$) such that
$$e^{-itH} = \int_\Omega F_t\,dM_{-i}^{t},\quad t\ge 0,$$
represents the dynamics of an interacting quantum field for some
selfadjoint operator $H$. {\it Any}\/ measurable multiplicative
functional $F_t:\Omega \to \bT$ is $M^t_\lambda$-integrable for
$\lambda > 0$. The existence of the dynamics for an
interacting field is determined by the {\it $M^t_\lambda$-integrability}\/
of $F_t$ in the sense of Section \ref{sec:Fpi}, for all $\lambda \in \overline{\C}_+\setminus\{0\}$ and $t > 0$.

According to \cite[pp. 132-133]{St}, the kernel $G_{\frac12}$ of the 
bounded linear operator $(-\Delta_{d-1}+1)^{-1/2}$
has the properties
\begin{eqnarray}
G_{\frac12}(x) &=& c|x|^{-(d-1)+1/2}+o(|x|^{-(d-1)+1/2})\quad \hbox{as } x\to 0,\label{eqn:7.14}\\
G_{\frac12}(x)&=& O(e^{-|x|/2})\quad \hbox{as } x\to \infty.\nonumber
\end{eqnarray}
Because 
$$\int_\Omega \langle f, X_t\rangle^2\,d\mu = 
\int_{\R^{2(d-1)}}f(x)G_{\frac12}(x-y)f(y)\,dxdy$$
for every $f\in \cS(\R^{d-1})$,
it follows from the estimate (\ref{eqn:7.14}) that
$\int_\Omega \langle f_n, X_t\rangle^2\,d\mu$ diverges as $f_n\to \delta_x$
weakly in the sense of measures as $n\to\infty$, for each $x\in \R^{d-1}$.
This is unfortunate, because polynomials in the random field are just the
type of interactions that need to be represented, for example, in the
quantisation of the classical $\phi^4$ field satisfying
$$-\square \phi +\phi + 4\gamma\phi^3 = 0$$
in Minkowski space \cite[p, 112]{GJ}.

For the purpose of discussing the most basic type of renormalisation
in the context of the construction of integrable multiplicative functionals
of a random field, attention is restricted to $d=2$. The
$1$-dimensional Laplacian $\Delta_1$ is actually the operator $d^2/dx^2$
with the domain of all functions $f\in L^2(\R)$
such that $\int_\R\xi^2|\hat f(\xi)|^2\,d\xi < \infty$.

For $n=1,2,\dots,$ let $\cP^{\le n}$ be the closed linear span in $\cH$
of all monomials 
$$\xi\longmapsto \prod_{j=1}^k\langle f_j,\xi\rangle,\quad \xi\in\cS'(\R),$$
for all $f_j\in \cS(\R)$, $j=1,\dots,k$ and $k=1,\dots,n$. The union
of all spaces $\cP^{\le n}$ is dense in $\cH$ and the Hilbert space
$\cH$ can be represented as the direct sum of the orthogonal complements
$\cP^{(n)} = \cP^{\le n}\ominus  \cP^{\le (n-1)}$ of $\cP^{\le (n-1)}$
in $\cP^{\le n}$ for $n=1,2,\dots$.

Then for each $f\in \cS'(\R)$ and $n=1,2,\dots$, there exists a unique
function $\Xi^n_f \in \cP^{(n)}$ such that
\begin{align*}
&\left(\Xi^n_f, \prod_{j=1}^n\langle f_j,\cdot\,\rangle\right)_\cH\\
 &\qquad=  
\frac{n!}{2^n}\int_\R\cdots\int_\R\prod_{j=1}^n
\left((-\Delta_1+1)^{-1/2}(y_j-x)f_j(y_j)\,dy_j\right)f(x)\,dx
\end{align*}
for all $f_1,\dots,f_n\in \cS(\R)$. In particular,
$f\longmapsto \Xi^n_f$, $f\in \cS(\R)$, is a linear
map from $\cS(\R)$ to $\cH$. The {\it Wick monomial} 
$\Xi^n_f$ is the orthogonal projection of the monomial function 
$\xi \longmapsto \langle f,\xi\rangle^n$, $\xi\in \cS'(\R)$ in $\cP^{(n)}$.

Denoting the Hermite polynomial of degree $n$
with leading coefficient one by $H_n$
and setting 
$$c(f)= \frac1{\sqrt2}\left\langle 
(-\Delta_{1} +I)^{-1/2}f,f\right\rangle^{1/2},$$ 
it follows \cite[Section 6.3]{GJ} that
$\Xi^n_f$ is the function 
$$\xi\longmapsto c(f)^{n}H_n(c(f)^{-1}\langle f,\xi\rangle),\quad
\xi\in \cS'(\R).$$
  Because $\cH$-valued function
$f\longmapsto \Xi^n_f$ is 
uniformly continuous with respect to the norm $f\longmapsto c(f)$ on $\cS(\R)$,
the random variable $\Xi^n_\Lambda$ may be defined by 
continuity as the element $\Xi^n_{\chi_\Lambda}$ of $\cH$ for every $\Lambda \in \cB_f(\R)$.

Our process $X$ takes values in $\cS'(\R)$, so the random variable 
$\Xi^n_\Lambda\circ X_t$ over the probability space
$(\Omega,\cF,\mu)$ makes sense and is traditionally written as
$$\int_{\Lambda} {:X^n_t:}(x)\,dx.$$
The mapping 
$$\Lambda \longmapsto \int_{\Lambda} {:X^n_t:}(x)\,dx,\quad \Lambda \in \cB_f(\R),$$
is an $L^2(\mu)$-valued measure but ${:X^n_t:}(x)$ is not defined as a random
variable, that is, there is no $L^2(\mu)$-valued density of the vector measure with respect to
Lebesgue measure on $\R$. 

A similar phenomena occurs with Brownian motion. If $m:\cB([0,T])\to L^2(\bP)$
is a mean zero Gaussian random measure with 
$$\bE(m(A)m(B))= |A\cap B|, \quad A,B\in \cB([0,T]),$$
then the vector measure $m$ has no $L^2(\bP)$-valued density with respect
to Lebesgue measure $|\cdot|$ on $[0,T]$. The density is distribution valued \textit{white noise}.

Nevertheless, the multiplicative functional
\begin{equation}\label{eqn:7.15}
F^{(\Lambda)}_t(\omega):= \exp\bigg[-i\int_0^t\left(\int_\Lambda
:X_s^{n}:(x)\, dx\right)(\omega) ds\bigg],\quad \Lambda \in \cB_f(\R),\ t\ge 0,
\end{equation}
has the property that $F^{(\Lambda)}_t$ is measurable with respect to
the $\s$-algebra generated by all random variables $\Phi_A$
with $A\in\cB(\R^2)$ and  $A\subset
\Lambda\times [0,t]$, $t\ge0$. It is also possible to
express the multiplicative functional (\ref{eqn:7.15}) 
as the limit of multiplicative functionals
of regularised processes $X^{(\e)}$ defined at points of $\R$ \cite[Proposition 8.5.1]{GJ}.

It is clear that the function $F^{(\Lambda)}_t$ is $M^t_\lambda$-integrable for
each $\lambda > 0$ because it is $M^t_\lambda$-measurable with absolute value one.
That $F^{(\Lambda)}_t$ is $M^t_\lambda$-integrable for $\lambda \in 
\overline{\C}_+\setminus\{0\}$ and $t\ge 0$ follows from the fact that
the closure of the operator
$-\left(\frac{1}{\lambda}H_0 +iQ(\Xi^n_\Lambda)\right)$ is the generator of a contraction
semigroup on $\cH$ for every $\lambda \in \overline{\C}_+\setminus\{0\}$,
see \cite[Section 5.3]{AHK1}.

The treatment of the $P(\phi)_2$ Euclidean fields in \cite{GJ}
is not sufficent to determine the limiting behaviour of the multiplicative functionals
$F^{(\Lambda)}_t$ as $\Lambda \nearrow \R$ simply because $F^{(\Lambda)}_t$
is a random variable with absolute value one for each $\Lambda \in \cB_f(\R)$.
The oscillatory nature of the expression (\ref{eqn:7.15}) is a new feature
and needs to be taken into account in order to make sense of 
formula (\ref{eqn:7.1}). 

By working in Minkowski space, rather than Euclidean space,
we may appeal to the interference between particle phases essential to Feynman's
physical reasoning.
Understanding the limiting process may lead to
new ways of constructing integrable multiplicative functionals $F$ of
quantum fields, and so giving the dynamical group $t\longmapsto e^{-itH}$, $t\in\R$, of the
associated interacting field determined by formula (\ref{eqn:7.1}).

\end{document}